\documentclass[12pt,preprint]{aastex}
\usepackage{natbib}
\usepackage{longtable}
\usepackage{color}
\definecolor{orange}{cmyk}{0,0.4,0.8,0.2}
\definecolor{darkorange}{rgb}{.71,0.21,0.01}
\definecolor{darkgreen}{rgb}{.12,.54,.11}

\usepackage{hyperref}
\hypersetup{pdftex,  
  breaklinks=true,  
  colorlinks=true,
  urlcolor=blue,
  linkcolor=darkorange,
  citecolor=darkgreen,
  }

\newcommand{\beq}{\begin{equation}}
\newcommand{\eeq}{\end{equation}}

\def\hipp {{\it Hipparcos}}
\begin{document}

\shorttitle{Mid-infrared Period--Luminosity Relations of RR Lyrae}

\shortauthors{C. R. Klein et al.}
\title{Mid-infrared Period--Luminosity Relations of RR Lyrae Stars Derived from the WISE Preliminary Data Release}
\author{Christopher R. Klein, Joseph W. Richards\altaffilmark{1}, Nathaniel R. Butler\altaffilmark{2}, and Joshua S. Bloom}
\affil{Astronomy Department, University of California, Berkeley, CA 94720;\\cklein@astro.berkeley.edu, 	jwrichar@stat.berkeley.edu, nat@astro.berkeley.edu, jbloom@astro.berkeley.edu}

\altaffiltext{1}{Statistics Department, University of California, Berkeley, CA 94720}
\altaffiltext{2}{NASA Einstein Fellow}

\slugcomment{Accepted for publication in ApJ, June 27th, 2011.}

\begin{abstract}
Interstellar dust presents a significant challenge to extending parallax-determined distances of optically observed pulsational variables to larger volumes. Distance ladder work at mid-infrared wavebands, where dust effects are negligible and metallicity correlations are minimized, have been largely focused on few-epoch Cepheid studies. Here we present the first determination of mid-infrared period-luminosity (PL) relations of RR Lyrae stars from phase-resolved imaging using the preliminary data release of the Wide-Field Infrared Survey Explorer (WISE). We present a novel statistical framework to predict posterior distances of 76 well-observed RR Lyrae that uses the optically constructed prior distance moduli while simultaneously imposing a power-law PL relation to WISE-determined mean magnitudes. We find that the absolute magnitude in the bluest WISE filter is $M_{W1} = (-0.421\pm0.014) - (1.681\pm0.147) \log_{10}(P/0.50118{\rm~day}),$ with no evidence for a correlation with metallicity. Combining the results from the three bluest WISE filters, we find that a typical star in our sample has a distance measurement uncertainty of 0.97\% (statistical) plus 1.17\% (systematic).  We do not fundamentalize the periods of RRc stars to improve their fit to the relations. Taking the \hipp-derived mean $V$-band magnitudes, we use the distance posteriors to determine a new optical metallicity-luminosity relation which we present in \S\ref{plrs_results}.  The results of this analysis will soon be tested by HST parallax measurements and, eventually, with the Gaia astrometric mission.


\end{abstract}

\subjectheadings{infrared: stars -- methods: statistical -- RR Lyrae: distance scale}

\section{Introduction}

RR Lyrae (RRL) pulsating variable stars are standardizable distance indicators at optical and near-infrared wavebands. In $V$-band, their brightnesses are nearly standard, with a small metallicity dependence and deviation about $\langle M_V \rangle$ of $\sim$0.12--0.15 mag \citep{1986ApJ...302..626H, 1998AA...330..515F,postHipp..book,2006ARA&A..44...93S}. At near-infrared wavebands RRL brightnesses are a well-fit function of pulsation period, with an apparently negligible  metallicity dependence (at $K$-band) and mean scatter from a period-luminosity (PL)  relation of $\sim$0.15 mag \citep{1986MNRAS.220..279L, 2006MNRAS.372.1675S}. The ability to infer distance to an RRL is chiefly restricted by the confidence in these empirically derived luminosity-metallicity and PL relations.

There is good observational and theoretical motivation to believe that infrared photometry offers the ability to derive more tightly constrained PL relations for pulsational variable stars in general. It has been argued \citep{1998salg.conf..263M} and demonstrated \citep{2008ApJ...679...71F, 2008MNRAS.386.2115F} that the scatter in these empirical relations is decreased at infrared wavelengths. \citet{1998salg.conf..263M} cite the advantages: (1) The sensitivity of surface brightness to temperature is a steeply declining function of wavelength; (2) The interstellar extinction curve decreases as a function of increasing wavelength (being almost linear with $1/\lambda$ at optical and near-infrared wavelengths); (3) At the temperatures typical of horizontal-branch stars, metallicity effects predominate in the UV, blue, and visual parts of the spectrum, where most of the line transitions occur, with declining effects at longer wavelengths. The overall insensitivity of infrared magnitudes of  RRL, Cepheid, and Mira variables to each of these effects results in decreased amplitudes for individual pulsating variables, as well as a decreased scatter in the apparent PL relations.

In this paper we present the first published mid-infrared PL relations for RRL variables. This is the first such work primarily because the requisite observational data has not existed previously. Since the farther reach of (brighter) Cepheid PL relations makes their study potentially more influential, the {\it Spitzer Space Telescope} has been used to derive mid-infrared PL relations for Galactic \citep{2010ApJ...709..120M} and  Magellanic Clouds \citep{2009ApJ...695..988M} Cepheids (the latter making use of SAGE survey data; \citealt{2006AJ....132.2268M,2009ApJ...695..988M}). These studies of Galactic (Magellanic Clouds) Cepheid mid-infrared PL relations reported best-fit dispersions of $\sim$0.2 mag ($\sim$0.12 mag).

RRL variables are particularly important local distance indicators because they are more numerous than Cepheids, and are observable within the Galactic disk and halo, within Galactic and some extragalactic globular clusters, and in the halos of neighboring dwarf galaxies (most notably, the LMC). Importantly, RRL variables can be used as stellar density tracers (e.g., \citealt{2001ApJ...554L..33V, 2010ApJ...708..717S}) to map the structure of stellar distributions.

In this article we derive mid-infrared RRL PL relations by analyzing observations of 76 RRL-type stars conducted with the Wide-field Infrared Survey Explorer (WISE) satellite \citep{2010AJ....140.1868W} and made available through the preliminary data release of the first 105 days of science data\footnote{\url{http://wise2.ipac.caltech.edu/docs/release/prelim/}}. We use a modified Lomb-Scargle \citep{lomb76,barning63,scargle82} period-finding algorithm to calculate the pulsation periods from both the WISE data and the very well-observed \hipp\ light curves of the same sources. We derive mean flux-weighted WISE magnitudes from the best-fit harmonic models of this Lomb-Scargle analysis; these observed magnitudes, along with the \hipp\ estimated periods, are used to estimate the WISE PL relations. The actual PL fitting is conducted through a Bayesian approach using {\it a priori} distance information and simultaneously fits the W1, W2, and W3 PL relations.  Our methods have general applicability, and can be used to robustly fit PL relations at other spectral wavebands. Our resulting mid-infrared PL relations are tightly constrained with absolute magnitude prediction uncertainties as small as 0.016, 0.016, and 0.076 mag at 3.4, 4.6, and 12 $\mu$m, respectively. 

The paper is outlined as follows.  We present a brief description of the WISE and ancillary data in \S\ref{data_desc}, followed by an explanation of our analysis methods in \S\ref{meth}. (In \S\ref{p_rec} we demonstrate the viability of period recovery with WISE data, and highlight the potential for discovery of new RRL variables and other short-period variables with the WISE single exposure database.) We describe the Bayesian method of deriving  mid-infrared PL relations in \S\ref{plrs} and discuss the results in \S\ref{plrs_results}. Finally, we present conclusions in \S\ref{concs}.

\section{Data Description}\label{data_desc}

WISE has imaging capabilities in four mid-infrared bands: W1 centered at 3.4 $\mu$m, W2 at 4.6 $\mu$m, W3 at 12 $\mu$m, and W4 at 22 $\mu$m. The satellite is in a polar orbit and scans the sky in great circles with a center located at the Sun and with a precession rate of 180$^\circ$ every six months \citep{2010AJ....140.1868W}. WISE completes about 15 orbits a day and the field of view of the detectors is 47 arcmin on a side. This configuration  allowed WISE to scan the entire sky in six months, with a minimum of 8 (median 12) single-frame exposures. Sources near ecliptic poles receive the most repeat coverage in time and sources near the ecliptic have the smallest number of observed epochs. WISE was launched on 2009 December 14 and operated until 2011 February 17. This mission duration provided two full scans of the sky. However, the hydrogen coolant ran out in 2010 October, halting data acquisition in the W3 and W4 bands.

The present study was conducted using data from the single-exposure database of the WISE preliminary data release, which was made public on 2011 April 14 through the NASA/IPAC Infrared Science Archive\footnote{\url{http://irsa.ipac.caltech.edu/}}. The preliminary data release includes the first 105 days of mission data and covers about 57\% of the sky.

The catalog of RRL variables used in the present study is derived from work by \citet{1998AA...330..515F}. The catalog contains 144 relatively local ($\leq$2.5 kpc) RRL variables selected from the \hipp\ catalog \citep{1997ESASP1200.....P} with color excess and metallicity measurements from previous literature.  Of the 144 RRL variables in our starting catalog, 77 were associated with sources in the WISE preliminary data release. We reject one light curve (V*EZLyr) because the reported WISE photometry does not indicate a periodic source (using the \hipp\ period; the source is also a strong outlier from our PL relation fits). All target sources, save the prototype RRL itself (V*RRLyr), are too faint for WISE to produce reliable W4 photometry and so we must ignore the longest wavelength data in the subsequent analysis.

To perform a proper analysis of the mid-infrared PL relation, in addition to the periods and observed WISE magnitudes of each RRL, we also need a prior guess of the distance to each object.  Here, we describe how we determine a prior distribution on the distance modulus of each RRL.  First, we compute the \hipp\ periods and mean flux-weighted magnitudes ($m_{\rm hip}$) using the same Lomb-Scargle based methods that we apply to the WISE light curve data (see \S\ref{meth}). Discrepancies with the published periods of \citet{1998AA...330..515F} were minimal (see Appendix \ref{p_rec}).  Unlike \citet{2006MNRAS.372.1675S}, we do not find that the periods of RRc type RRL variables need to be fundamentalized by adding a constant term ($\Delta \log_{10} P \approx 0.13$) in order to improve the PL relation scatter.  Following  \citet{1998ApJ...508..844G} we determine values of the apparent Johnson $V$-band magnitude ($m_V$) and the effective extinction, $A_{\rm V, eff}$, that differ slightly from  \citet{1998AA...330..515F}.  This is achieved by making use of the line-of-sight extinction from the  \citet{1998ApJ...500..525S} (SFD) dust maps and by assuming a Galactic scale-height model for the dust (such that not all SFD dust lies in between us and the RRL). In particular, we determine an effective extinction for the $i$th RRL in the sample as:
\begin{equation}
 E(B-V)_{\rm eff,i} = E(B-V)_{\rm SFD,i} \left( 1 - \exp[-|z_i|/h]\right),
\end{equation}
where $E(B-V)_{\rm SFD,i}$ is the differential SFD extinction towards source $i$, $z_i$ is the scale height of $i$th source above the Galactic plane\footnote{Note that $z_i$ will depend on the value of distance determined, however, we simply use the coordinate information provided by \citet{2005A&A...442..229M} when available or transform the sky coordinates using the \citet{1998AA...330..515F} distance results when necessary. Our results are not sensitive to the precise value of $z_i$.}, and $h=130$ pc is the exponential scale height assumed for the dust in the disk \citep{1998ApJ...508..844G}. To convert  $m_{\rm hip}$ to $m_V$ we adopt the prescription from  \citet{1998ApJ...508..844G}:
\begin{equation}
m_{V,i,{\rm eff}} = m_{\rm hip,i} - X - 0.2 E(B-V)_{\rm eff,i}
\end{equation}
where $X=0.09$ for RRab types and $X=0.06$ for RRc types. We assume a 15\% error on $E(B-V)_{\rm eff,i}$. Finally, we produce an extinction corrected magnitude $m^*_{V,i} = m_{V,i,{\rm eff}} - A_{V, {\rm eff}, i}$, where $A_{V, {\rm eff}, i} = 3.1 \times E(B-V)_{{\rm eff},i}$, and the factor $R=A_V/E(B-V)=3.1$ from \citet{1975A&A....43..133S}. These values of extinction-corrected $V$-band magnitudes (and associated errors) are reported in Table \ref{inputs_table}. 

To determine the prior distance modulus, $\mu_{0,i}$, for the $i$th source, we need a prescription for determining the absolute $V$-band magnitude given the metallicity of the star (there is no known relationship of period and luminosity at $V$-band for RRL variables). We adopt the $M_V$--metallicity relation given in \citet{postHipp..book}, where we use the metallicity data as provided in \cite{1998AA...330..515F}. Explicitly, the $M_V-$[Fe/H] relation used is 
\begin{equation}
\label{eq:Mv-Z}
M_V = (0.23 \pm 0.04)([{\rm Fe}/{\rm H}]+1.6) + (0.56 \pm 0.12). 
\end{equation}
The calculated values of $M_{V,i}$ for each source are given in Table \ref{inputs_table}. 

Finally, we compute the prior mean of the distance modulus of the $i$th RRL as $\mu_{0,i} = m^*_{V,i} - M_{V,i}$, with the uncertainty in this quantity propagated assuming the errors on $m^*_{V,i}$ and $M_{V,i}$ are Gaussian and uncorrelated. The values of $\mu_{0,i}$ (and $\sigma_{\mu_{0,i}}$) (Table \ref{inputs_table}) represent our best estimates of the distances (and errors) using the body of work on RRL variables at visual bands \emph{prior} to analyzing the WISE data and \emph{prior} to the {\it Hubble Space Telescope} (HST) parallax result on V*RRLyrae itself. Note however that our prior estimate on distance modulus to V*RRLyrae ($\mu_{0} = 7.042 \pm 0.125$) is consistent with that found directly from HST parallax measurements ($\mu = 7.090 \pm 0.063$; \citealt{2002AJ....123..473B}).

\section{Light Curve Analysis Methods}\label{meth}

\begin{figure}[p]
\centerline{\includegraphics[width=4in]{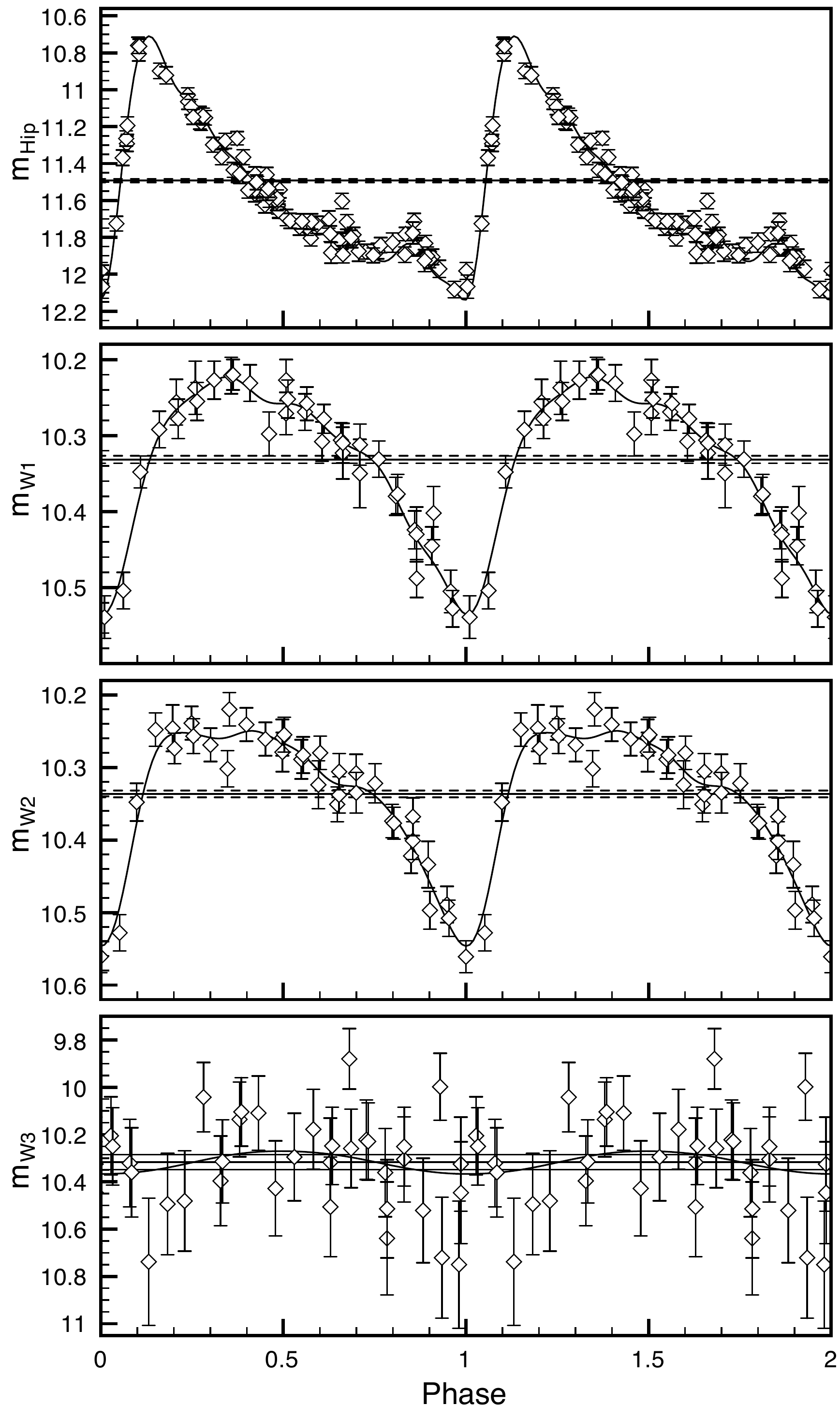}}
\caption{\hipp\ and WISE light curves of V*UPic with fitted models. The solid horizontal line is at the mean flux magnitude and the dashed lines represent the uncertainty of this mean magnitude. The model fitted to the \hipp\ (WISE W1) data uses 8 (7) harmonics. V*UPic is our best observed WISE source with 35 observations over $\sim$2.3 days. The data is phase-shifted so that the minimum of the model in each band is at ${\rm phase}=1$.}
\label{UPic_light_curve}
\end{figure}

In order to determine the PL relations for WISE, we need to calculate an apparent brightness. Following common practice \citep{1998AA...330..515F, 1990ApJ...354..273L}, we define the brightness of each source as the mean flux,
converted to a magnitude.  As we expect possible poor phase sampling
in the WISE data, we use a model --- based on a modified Lomb-Scargle
algorithm \citep{2011arXiv1101.1959R}, which allows for data uncertainty
and a mean flux offset --- instead of the observed data
points, to determine this mean (see Figure \ref{UPic_light_curve} for an example).  At significant peaks in the periodogram,
this model construction attempts to fit as many as 8 harmonic
components --- at frequencies which are multiples of the fundamental frequency
--- in addition to the fundamental frequency component.  Complex models
are penalized using generalized cross validation (e.g., \citealt{SSS2009.HTF}) to prevent over-fitting.  The resulting model
curves are smooth, typically
dominated by the presence of 4--6 harmonics for \hipp\ data, and
can be used to
calculate the flux integral and its uncertainty.  For the case of \hipp,
we find that the difference between our flux estimates and those from \cite{1998AA...330..515F} exhibit an rms scatter of 1.4\% with no systematic difference.

Applied directly to the WISE data, our period finding framework accurately
recovers the majority of the RRL periods directly from the
WISE data (see Appendix \ref{p_rec}).  For all WISE mean-magnitude estimates, we force the Lomb-Scargle model to use the best-fit \hipp\ periods as the fundamental frequency.
We note that our mean-magnitude estimates remain unchanged, within their
uncertainties, if we instead use the best-fit WISE periods.

\section{Deriving the PL Relations}\label{plrs}

Using the full sample of 76 WISE RRL variables, we derive 
the empirical PL relationship in each of the bands W1, W2, and W3.  For each RRL in the sample,
we estimate the observed magnitude, $m_{ij}$, and pulsational period, $P_i$, using the methods outlined in \S\ref{meth}.
Here, $i=1,...,n$ indexes the RRL variables and $j=1,2,3$ indexes the WISE bands.

Our statistical model of the PL relationship is\footnote{In principle, there could be a metallicity dependence, but we found such a dependence was negligible in the WISE bands. See \S \ref{plrs_results}.}
\begin{equation}
\label{eq:model}
m_{ij} = \mu_i + M_{0,j} + \alpha_j \log_{10}(P_i/P_0) + \epsilon_{ij},
\end{equation}
where $\mu_i$ is the distance modulus for $i$th RRL, $M_{0,j}$ is the absolute magnitude zero point for the $j$th WISE
band at $P=P_0$, where $P_0 = 0.50118$ day is the mean period of the sample, and $\alpha_j$ is the slope of the PL 
relationship in the $j$th band.  We assume that any extinction is negligible in these bands. The error terms $\epsilon_{ij}$
are independent zero-mean Gaussian random deviates with variance $(\sigma \sigma_{m_{ij}})^2$, which describe the intrinsic scatter in the $m_{ij}$ about the model, where $\sigma$ is a free parameter which is an unknown scale factor on the known measurement errors, $\sigma_{m_{ij}}$\footnote{The average measurement error, $\sigma_m$, is 0.013, 0.013, and 0.045 mag in W1, W2, and W3, respectively.}. We fit the model (eq.~\ref{eq:model}) using a 
Bayesian procedure, described below.

A Bayesian approach to this problem is appropriate because for each RRL we have {\it a priori} distance information from previous V-band RRL studies.  For each RRL in our sample, we determine a prior on its distance modulus using the steps outlined in \S\ref{data_desc}.  For the star V*RRLyr, we adopt the HST distance estimate of \cite{2002AJ....123..473B} as our prior.  These priors encompass the full amount of information that we have about each source's distance \emph{before} looking at the WISE data.
The key in our analysis is that while the distance to any RRL could be changed within its  prior to fit a perfect PL relation in a single band, the simultaneous fitting of a power-law PL relation in all bands (with as little intrinsic scatter as possible) tightly constrains the distance of each source.
Bayesian fitting of the PL model allows us to obtain:
\begin{itemize}
\item posterior distributions on the distance to each RRL, given the WISE data,
\item posterior distributions on the absolute magnitude zero point and slope of the PL relationship in each WISE band, and
\item an estimate of the amount of intrinsic spread of the data around the PL relationship.
\end{itemize}
The end goal, of course, is to use the estimated PL relationship to accurately \emph{predict} the distance to each newly observed RRL from its period and observed WISE light curve.  Furthermore, we want to make these predictions with an accurate notion of the amount of error in each predicted distance, as those errors will propagate to subsequent studies.

Bayesian fitting of  linear models is thoroughly described in \cite{gelman2003}.  Here, we summarize our procedure 
for analysis of the WISE PL relationship.  First, we assume a normal (Gaussian) prior distribution on
each of the distance moduli with mean $\mu_{0,i}$ and standard deviation $\sigma_{\mu_{0,i}}$, as described above.
For the other parameters in our model (eq.~\ref{eq:model}), we assume a
flat, noninformative prior distribution.  For convenience, we rewrite the model in matrix form as $m = 
\mathbf{X}\beta + \epsilon$, where $m$ is a vector of the $3n$ measured WISE mean-magnitudes,
$\beta$ is a vector of the $n+6$ parameters ($\mu, M_0, \alpha$) in the PL model, $\mathbf{X}$ is the
appropriate $3n$ by $n+6$ design matrix for eq.~\ref{eq:model}, and $\epsilon$ is a vector of the zero-mean, normally
distributed random errors with covariance matrix $\sigma^2 \textrm{diag}(\sigma_m^2)$.

Including an informative prior on $\mu$ is equivalent to adding extra prior ``data points'' to the analysis.  In our model,
these ``data points'' are $\mu_{0,i} = \mu_i + \sigma_{\mu_{0,i}}\epsilon_i$, where $\epsilon_i$ is a  normal random variate
with mean 0 and variance 1.
This prior information on $\mu$ induces the model $m_* = \mathbf{X}_*\mathbf{\beta} + \epsilon_*$, where
\begin{eqnarray}
\label{eq:lmbayes}
m_* &=& \left( \begin{array}{c}
        m\\
        \mu_0 \end{array} \right), \hspace{.05in}\\
  \mathbf{X}_* &=& \left( \begin{array}{c}
        \mathbf{X}\\
        (I_n, 0_{n,6} ) \end{array} \right),      \hspace{.05in}\\
\label{eq:lmsig}
 \Sigma_* &=& \left( \begin{array}{cc}
        \sigma^2 \textrm{diag}(\sigma_m^2) & 0_{3n,n}\\
        0_{n,3n} & \textrm{diag}(\sigma_{\mu_0}^2)\end{array} \right),
\end{eqnarray}
 $\epsilon_* \sim N(0,\Sigma_*)$, and $N$ denotes the multivariate normal distribution.
Here, $I_n$ indicates the $n\times n$ identity matrix and $0_{m,n}$ is the $m\times n$ matrix of 0s.

Posterior distributions for the parameters of interest can be derived in a straightforward manner using the  
entities in eqs.~\ref{eq:lmbayes}--\ref{eq:lmsig}.  The joint posterior distribution, $P(\beta,\sigma | m,P)$, can be sampled by
first drawing from $P(\sigma | m,P)$ and then, conditional on that draw, selecting from $P(\beta | m,P,\sigma)$.
 The posterior distribution for
 $\beta$, conditional on the value of $\sigma$, follows the multivariate normal distribution,
\begin{equation}
\label{eq:betapost}
\beta | m,P,\sigma \sim N(\widehat{\beta}, (\mathbf{X}_*'\Sigma_*^{-1}\mathbf{X}_*)^{-1})
\end{equation}
where $\widehat{\beta}$ is the standard maximum likelihood (weighted least squares) solution,
\begin{equation}
\widehat{\beta} = (\mathbf{X}_*'\Sigma_*^{-1}\mathbf{X}_*)^{-1}\mathbf{X}_*'\Sigma_*^{-1}m_*.
\end{equation}
Unlike the posterior distribution of $\beta$ (given $\sigma$), the posterior distribution of $\sigma$ does not follow
a simple conjugate distribution.  Instead, the distribution  follows the form
\begin{equation}
\label{eq:sigpost}
P(\sigma^2 | m,P) \propto \frac{P(\beta)P(\sigma^2) L(m | P,\beta,\sigma)}{P(\beta | m,P,\sigma)}
\end{equation}
where the prior on $\beta$ is proportional to the informative prior on $\mu$, the flat prior on $\sigma$ is $P(\sigma^2) \propto \sigma^{-2}$, and the data likelihood $L$
is the product, over all  observed magnitudes, of the Gaussian likelihood of the data given the model (eq.~\ref{eq:model}) with all parameters specified.

We  draw samples from our joint posterior distribution  $P(\beta,\sigma | m,P)$ using eqs.~\ref{eq:betapost} and 
\ref{eq:sigpost} in conjunction.  In practice, we compute\footnote{Assuming
 that $\beta = \widehat{\beta}$.  Several iterations show that the posterior distribution of $\sigma$
is insensitive to the assumed choice of $\beta$.} $P(\sigma^2|m,P)$ over a fine grid of $\sigma$ values using eq.~\ref{eq:sigpost}, and then draw a sample of $\sigma$  from this density.
For each sampled $\sigma$, we subsequently draw a $\beta$ from eq.~\ref{eq:betapost}, conditional on the drawn $\sigma$ value.
We repeat this process 10,000 times to characterize the joint posterior distribution.
Using a large sample from this joint posterior distribution, we can compute  quantities of interest such as the
maximum {\it a posteriori} slopes and zero points of the PL relationship of each WISE band, the intrinsic scatter of the data around
the PL relationship in each band, and the spread in the {\it a posteriori} distribution of the PL parameters (see Figs.~\ref{scatter_w1}--\ref{scatter_w3}).

\section{PL Relations Discussion}\label{plrs_results}

\begin{figure}[p]
\centerline{\includegraphics[width=4in]{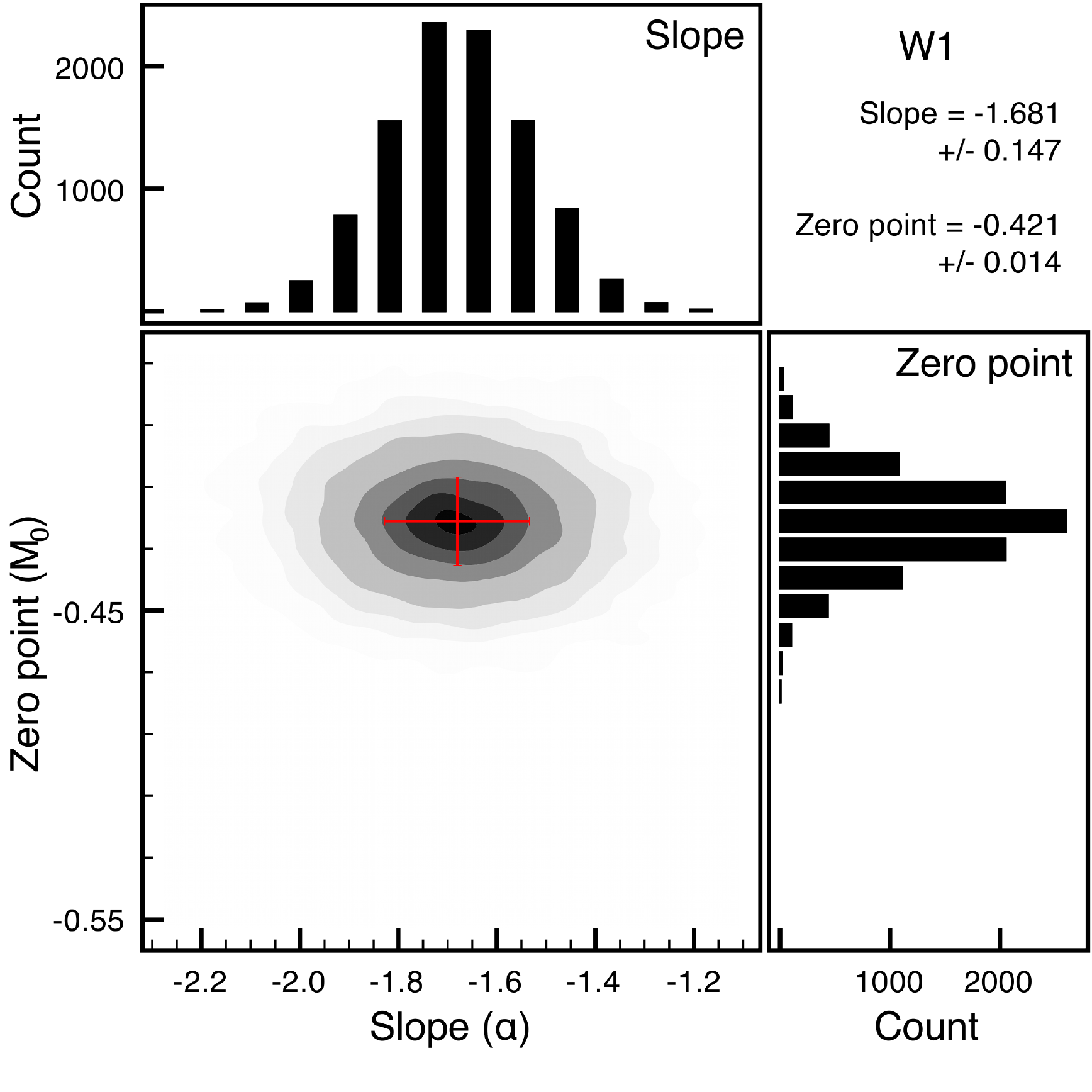}}
\caption{Contour plot and histograms of 10,000 samples from the posterior distribution of the slope ($\alpha_1$) and absolute magnitude zero point ($M_{0,1}$) of the period-luminosity relation for W1.  Our data constrain $\alpha_1$ to $-1.681 \pm 0.147$ and $M_{0,1}$ to $-0.421 \pm 0.014$, with negligible correlation between those parameters. Levels in the 2D contour plot are at the 99.9, 99, 97.5, 95, 90, 85, 80, and 70th percentile.}
\label{scatter_w1}
\end{figure}

\begin{figure}[p]
\centerline{\includegraphics[width=4in]{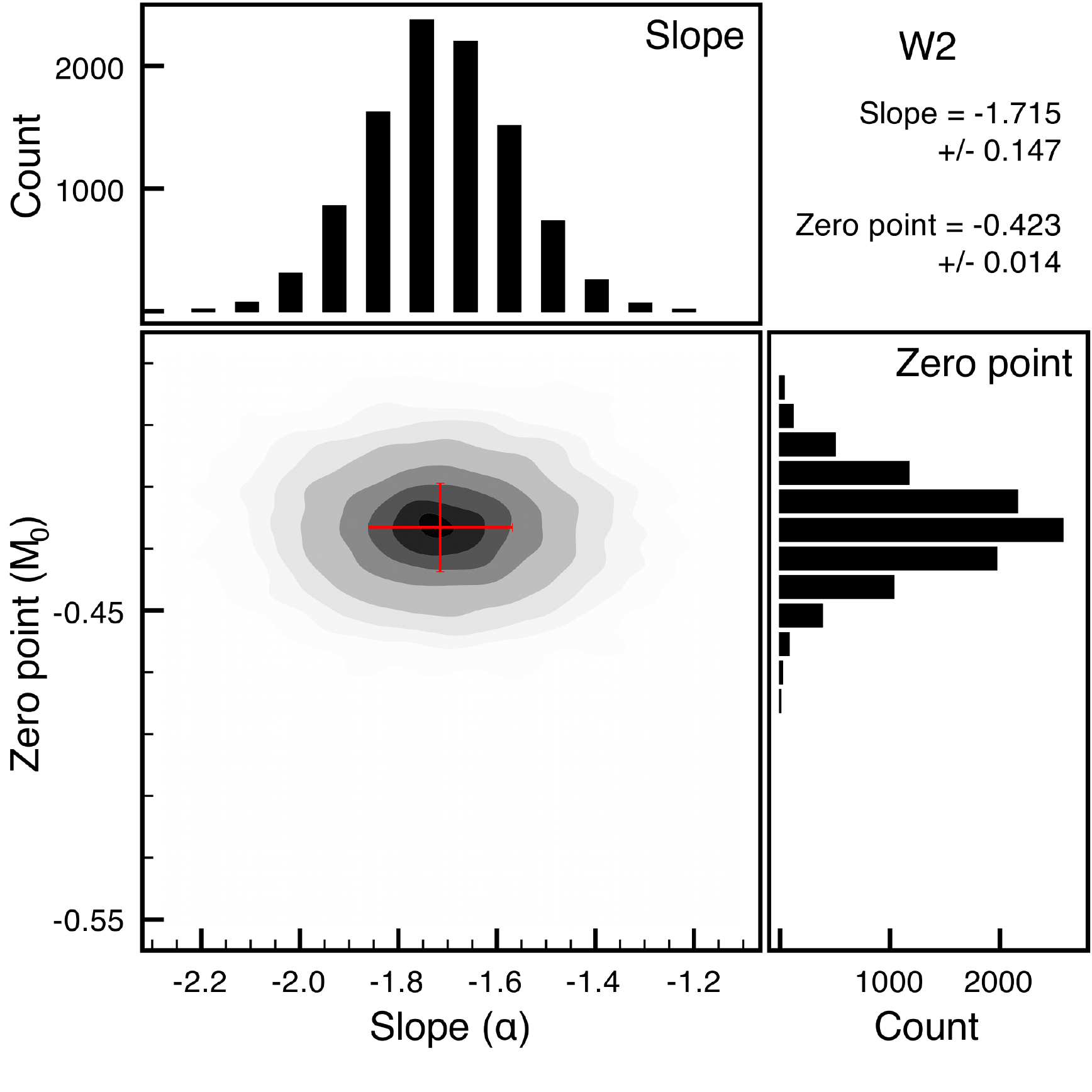}}
\caption{Same as in Figure \ref{scatter_w1}, for W2.  Our data constrain $\alpha_2$ to $-1.715 \pm 0.147$ and $M_{0,2}$ to $-0.423 \pm 0.014$, with negligible correlation between those parameters.}
\label{scatter_w2}
\end{figure}

\begin{figure}[p]
\centerline{\includegraphics[width=4in]{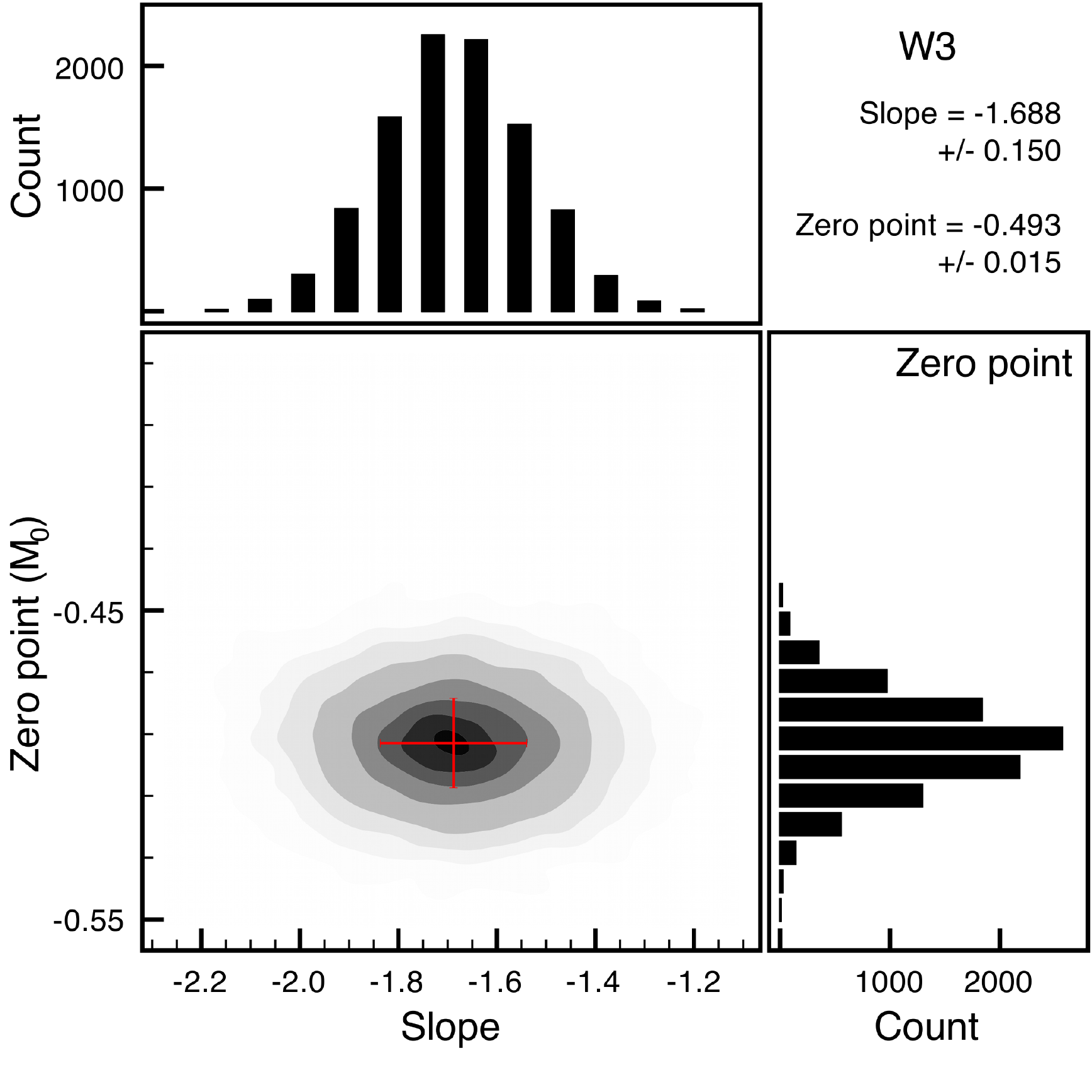}}
\caption{Same as in Figure \ref{scatter_w1}, for W3.  Our data constrain $\alpha_3$ to $-1.688 \pm 0.150$ and $M_{0,3}$ to $-0.493\pm 0.015$, with negligible correlation between those parameters.}
\label{scatter_w3}
\end{figure}

Bayesian analysis of the WISE RRL variables shows a strong PL relationship in each of the three bands.  The maximum
{\it a posteriori} estimates (and corresponding errors) of the slopes and absolute magnitude zero points for each of the three bands (eq.~\ref{eq:model}) and the joint posterior distributions of these parameters are plotted in
Figures \ref{scatter_w1}, \ref{scatter_w2}, and \ref{scatter_w3} (see also \S \ref{concs}).  
At the mean period ($P_0= 0.50118$ day) of the sample, we achieve an absolute magnitude prediction error of 0.016, 0.016, and 0.076 mag in W1, W2, and W3, respectively.  Therefore, for an RRL of period near 0.5 day observed in WISE W1 or W2 bands, we can predict the absolute magnitude of that object to within 0.016 mag, which corresponds to a fractional distance error of 0.7\%.

\begin{figure}[p]
\centerline{\includegraphics[width=6.55in]{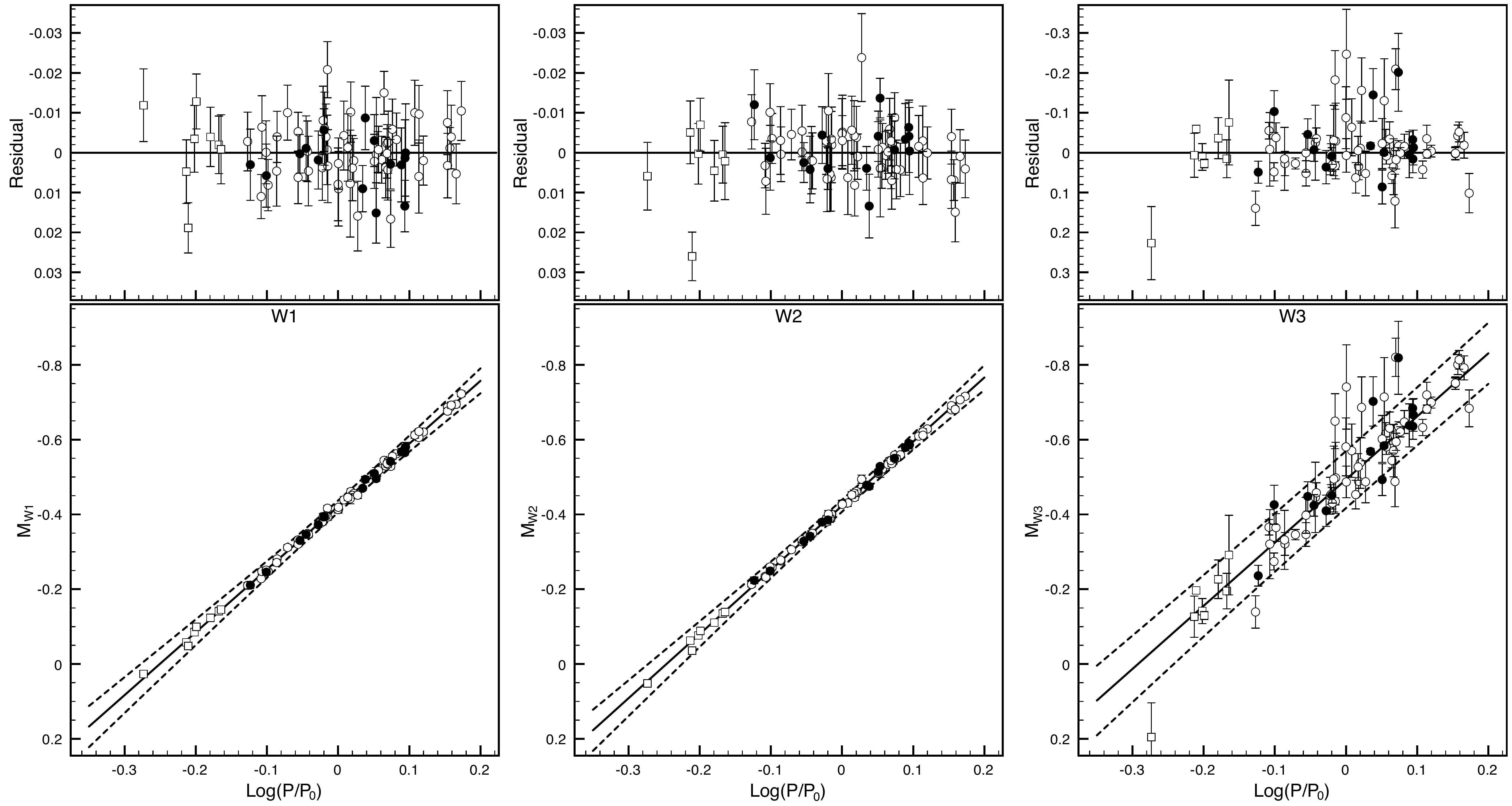}}
\caption{Period-luminosity relations for W1, W2, and W3 (left to right), as found by our Bayesian fitting method. In each figure, the solid line shows our model's prediction of the RRL absolute magnitude, as a function of RRL period.  The dashed lines show the $\pm 1\sigma$ prediction intervals; we expect the true absolute magnitude to fall within the dashed lines for 68\% of all newly observed RRL variables.  The top panel of each plot shows the residual spread around the best fit model, showing the small variance, $\sigma^2$, in the intrinsic scatter around the PLRs. In the figure, RRab are plotted as circles and RRc as squares.  Blazhko-affected RRL are indicated by filled points.}
\label{plr_fig}
\end{figure}

The width of the absolute magnitude prediction bands becomes slightly larger as one moves to larger or smaller periods, as the model is less constrained in those regions.  However, the prediction uncertainty remains low throughout the full period range, even at the extremes.  For example, at a period of 0.3 day, the absolute magnitude prediction error is 0.037, 0.037, and 0.084 mag (1.7, 1.7, and 3.9\% fractional distance error); at a period of 0.7 day, the prediction errors are 0.026, 0.026, and 0.079 mag (1.2, 1.2, and 3.6\% fractional distance error) for W1, W2, and W3, respectively.  Figure \ref{plr_fig} plots the estimated PL relationship in each WISE band, plus the $\pm 1\sigma$ prediction intervals.  For each newly observed RRL, the true absolute magnitude is expected to reside within the prediction interval.

\begin{figure}[p]
\centerline{\includegraphics[width=4in]{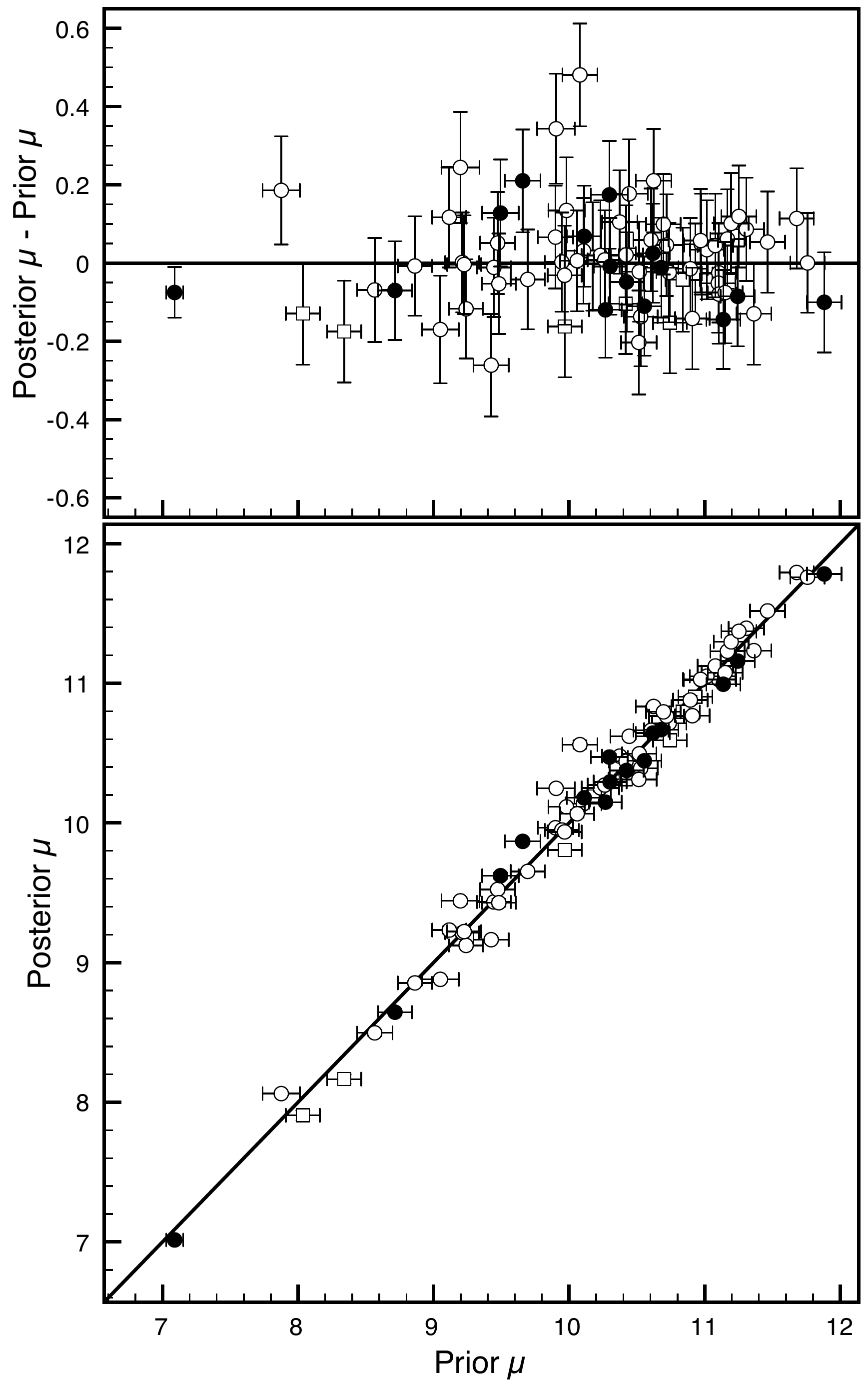}}
\caption{Prior versus posterior distance moduli (bottom) and residual difference (top). The posterior distance modulus distribution is determined by sampling from the density in eq.~\ref{eq:betapost}, which considers the evidence in all three WISE bands as well as the prior distance information.  As is evident, the posterior distances are consistent with the prior distance distribution, within their errors.  In the figure, RRab are plotted as circles and RRc as squares.  Blazhko-affected RRL are indicated by filled points.}
\label{mu_mu}
\end{figure}

Along with estimating the PL relationship for each band, our fitting procedure supplies a posterior distribution for the distances of each of the RRL in our sample.  In Table \ref{distances_table} we report the posterior means along with the 68\% and 95\% posterior credible sets for the distance to each of the 76 RRL used to fit the PL relationships. We also list the separation between prior and posterior distance moduli in units of $\sigma$, defined as 
$$\Delta({\mu}_{\rm prior}-{\mu}_{\rm post}) = \frac{\bar{\mu}_{\rm prior}-\bar{\mu}_{\rm post}}{\sqrt{\sigma_{\mu_{\rm prior}}^2 + \sigma_{\mu_{\rm post}}^2}},$$
where $\bar{\mu}$ and $\sigma_{\mu}$ denote the means and standard deviations of the distributions, respectively.  We note that there is, for most RRL variables, a close correspondence between the prior and posterior distances, as $|\Delta| < 2$ for all but 2 sources in our sample (V*ANSer and V*HKPup).   Figure \ref{mu_mu} shows a plot of prior versus posterior distance moduli, including a residuals plot, which shows again that, within their errors, the posterior distance distributions are consistent with the prior distance distributions for almost all the RRL.

\begin{figure}[p]
\centerline{\includegraphics[width=3.20in]{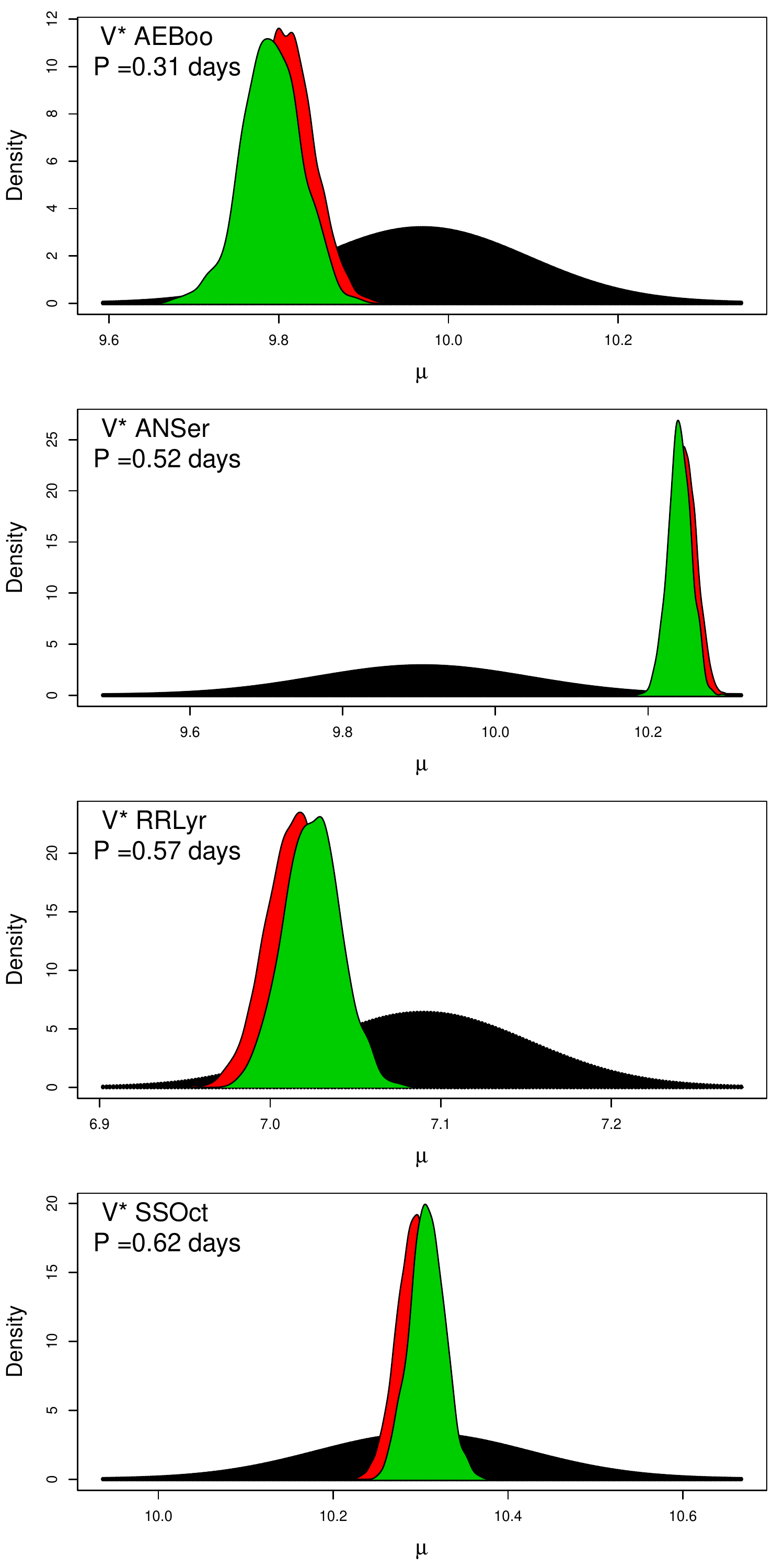}}
\caption{Comparison of the prior, posterior, and prediction density for the distance modulus, $\mu$, for 4 RRL in our sample.  From top to bottom, these sources are V*AEBoo, V*ANSer, V*RRLyr, and V*SSOct.  In each plot, the broad black curve represents the prior $\mu$ density, the red (hidden) curve is the posterior density of $\mu$, and the green (foreground) curve is the prediction density for $\mu$, which is found by holding that source out during model fitting and then predicting its $\mu$ with the built model.  For three of the four stars, the posterior densities are in good agreement with the prior; V*ANSer is the second most discrepant star in our sample (after V*HKPup).  In all cases, the posterior and predictive densities are much more precise than the prior densities and are in very close agreement to one another.}
\label{plr_fit_mu_comparison}
\end{figure}

Recall that for V*RRLyr we use the well-measured HST parallax result, which corresponds to $262 \pm 7.5$ pc. Our posterior fit distance for V*RRLyr is $253 \pm 2$ pc, which is consistent with the HST distance at a level of $1.2\sigma$. We also get a consistent prediction for V*RRLyr if we do not use V*RRLyr itself in the PL analysis (Fig.~\ref{plr_fit_mu_comparison}). That our analysis for the source with the most highly constrained distance prior is consistent with those results is further evidence of its accuracy and applicability.

To check the sensitivity of our results to the prior distances used, we analyze the changes in our posterior distance estimates under systematic prior offsets.  We first note that the prior $\mu$ estimate for V*RRLyr using the HST parallax result differs by 0.05 dex from the \hipp\ V-band estimate.  To estimate the amount of systematic error in our posterior distance estimates, we inflate the prior mean distance modulus by 0.05 dex for a random 50\% of the RRL before running our Bayesian PL model fitting.  As a result, the posterior distance moduli increase by an average of 0.023 dex, implying a systematic error of 1.17\% on distance estimation. We take this to be a reasonable estimate of the systematic error.
 
As a further sanity check, in Figure \ref{plr_fit_mu_comparison} we compare the prior, posterior, and prediction $\mu$ densities for a few RRL variables.  The prediction density for each RRL was computed by holding out that particular  RRL during the model fitting, and then applying the fitted model to predict the distance modulus of that source.  We find that these ``cross-validated'' $\mu$ prediction densities are very consistent with the posterior $\mu$ densities, suggesting that the model is stable and that small changes in the set of RRL used to fit the model do not cause any substantial differences in the model.  Furthermore, those densities are much more narrow than the prior densities, showing that the WISE data can constrain the distances to a great degree.  Additionally, we see that both the posterior and prediction densities fall within high-probability regions of the prior distribution for three of the four stars, meaning that our model is in good agreement with the prior distances.  Note that the one discrepant star plotted, V*ANSer, has the second largest discrepancy between prior and posterior $\mu$ densities, after V*HKPup (Table \ref{distances_table}).

We also test whether including RRL metallicity into the model improves the PL relationship fits.  To do this, we add an additional term, $\gamma_j Z_i$, to our model (\ref{eq:model}), where $Z_i$ is the metallicity of RRL $i$ and $\gamma_j$ is the slope of the magnitude-metallicity relationship for the $j$th WISE band.  Fitting this model, we find that $\gamma_j$ has a significantly positive value, but  that the predictive power of the new model, as measured by the width of the prediction intervals around the absolute magnitudes, does not differ from the original model which neglected metallicity.  Furthermore, if we first subtract from the absolute magnitudes the fit of the model that uses only period, we find no relationship between the residuals and  metallicity (slope of $-0.00034 \pm 0.00151$).  Including only period in the model achieves significantly better fits than including only metallicity, with half as much residual scatter.  These results suggest that all of the absolute magnitude information encoded in [Fe/H] is already contained in the period, and so metallicity need not be added as a covariate in the model.

\begin{figure}[p]
\centerline{\includegraphics[width=4in]{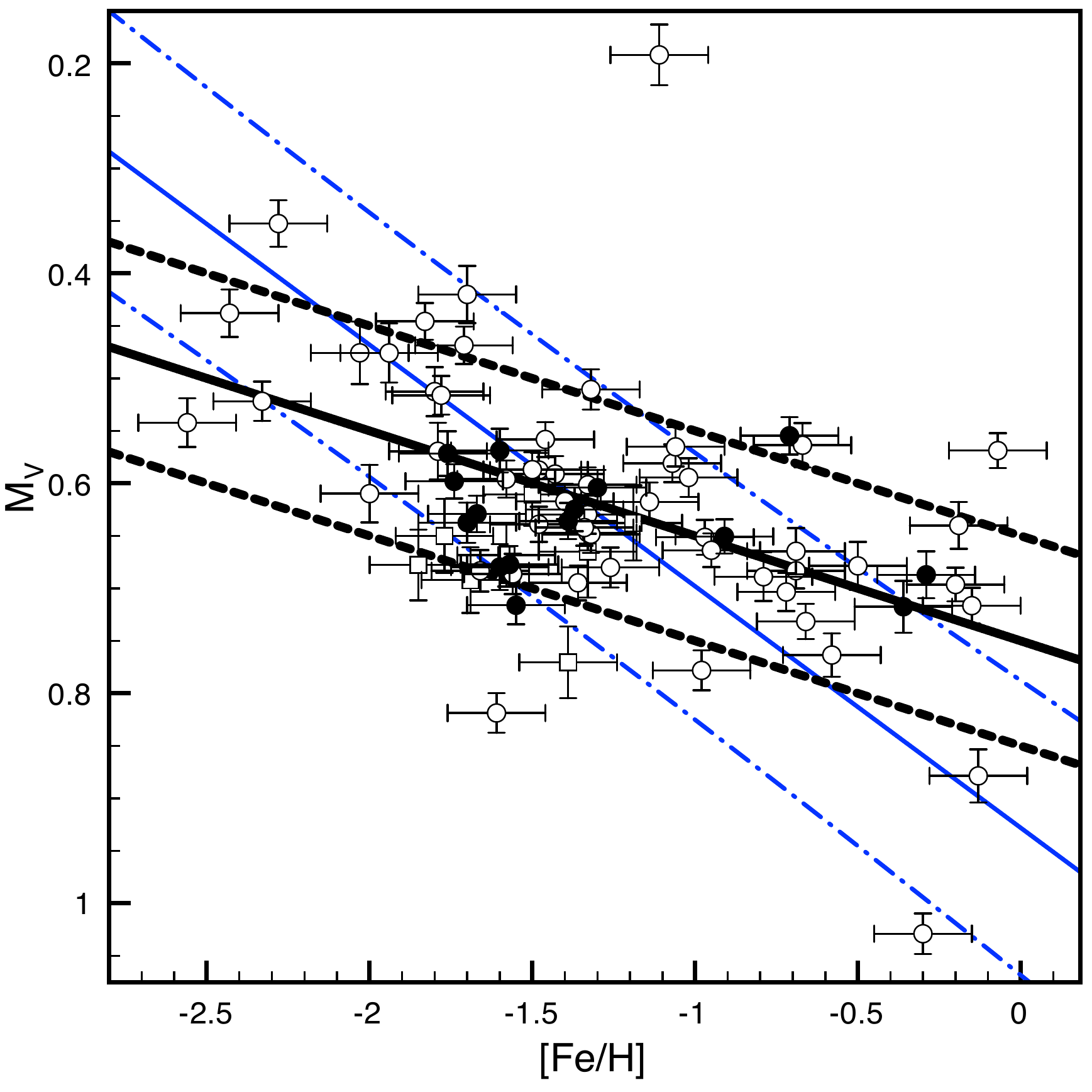}}
\caption{$M_V-$[Fe/H] relation as derived from our new, WISE-constrained RRL distances. In the figure, RRab are plotted as circles and RRc as squares. Blazhko-affected RRL are indicated by filled points. The black thick lines depict our fit of the relation, $M_V = 0.10([{\rm Fe}/{\rm H}]+1.6) + 0.59$, while the amount of intrinsic scatter about the fit is $\pm 0.10$ mag, represented by the black dashed lines.  The blue thinner lines depict the relation from \cite{postHipp..book}, $M_V = (0.23 \pm 0.04)([{\rm Fe}/{\rm H}]+1.6) + (0.56 \pm 0.12)$, with dash-dot lines showing the $\pm1\sigma$ bounds of the relation.}
\label{V_Z_relation}
\end{figure}

Finally, we derive an empirical $M_V-$[Fe/H] relationship using the posterior mean $\mu$ values from our Bayesian fitting to the WISE data.  From this data, our best fit relationship is $M_V = (0.10 \pm 0.02)([{\rm Fe}/{\rm H}]+1.6) + (0.59 \pm 0.10)$, which differs significantly in its slope, but not in its intercept value, to the \cite{postHipp..book} relationship---in eq. \ref{eq:Mv-Z}---that was used to compute the original distance priors.  Figure \ref{V_Z_relation} shows a scatterplot of our estimated $M_V$ as a function of metallicity; there is significant scatter around the empirical relationship, with a handful of large outliers. We also overplot the \cite{postHipp..book} relation to demonstrate that both relations fit the data reasonably well. We qualify our new $M_V-$[Fe/H] relation by noting that the relatively constrained metallicity range of our sample RRL variables limits the relation's applicability at other metallicities. As RRL metallicity deviates from $\sim$$-1.5$ the uncertainty in the slope of the relation rises steeply.

At a glance, the nontrivial difference between the  \cite{postHipp..book} $M_V-$[Fe/H] relationship used to calculate our distance priors and the new $M_V-$[Fe/H] relationship we derive using our distance posteriors could indicate an inconsistency in the Bayesian approach to our PL relation fits. In particular, this discrepancy may suggest that the large spread in the prior distance distributions has allowed the Bayesian fitting technique too much freedom in computing posterior distances. To test this, we run a simple weighted least squares regression to fit each of the PL relations, fixing the distances at the exact values from the  \cite{postHipp..book} $M_V-$[Fe/H] relationship (without using a Bayesian fitting method to update the distance estimates). This simpler fitting method results in statistically identical slope and zero point parameters for all three WISE bands. The scatter about the least squares fit, however, increases to 0.12 mag in W1 and W2, and 0.15 mag in W3 (from 0.016, and 0.076 from the Bayesian method). This increased scatter is expected, since the primary purpose of applying the Bayesian fitting technique is to reduce this scatter by simultaneously finding more accurate distances through the posterior distribution (i.e., updating the distance estimates given the WISE data). We can thus state confidently that the discrepancy between the  \cite{postHipp..book} $M_V-$[Fe/H] relationship and the new $M_V-$[Fe/H] relationship that we derive does not affect the PL relation fits.

\section{Conclusions} \label{concs}

We have presented the first  calibration of the RRL period-luminosity relations at three mid-infrared wavelengths.  Our estimated PL relations, tied to the Vega magnitude system, are:
\begin{eqnarray}
M_{W1} &=& (-0.421\pm0.014) - (1.681\pm0.147) \log_{10}(P/0.50118{\rm~day})\\
M_{W2} &=& (-0.423\pm0.014) - (1.715\pm0.147) \log_{10}(P/0.50118{\rm~day})\\
M_{W3} &=& (-0.493\pm0.015) - (1.688\pm0.150) \log_{10}(P/0.50118{\rm~day}).
\label{eq:ans}
\end{eqnarray}
\noindent These relations achieve an absolute magnitude prediction error as low as $\pm 0.016$ mag in WISE bands W1 and W2 (rising to $0.076$ mag in W3) near the mean period value $P_0=0.50118$ day. Using these relations we calculated new distances to our sample of RRL stars with a mean fractional distance error of 0.97\% (statistical) and 1.17\% (systematic). 

We further demonstrated that the posterior distances resulting from the newly-derived PL relations are consistent with the prior distance distributions. An attempt to find an independent, statistically significant metallicity dependence in the mid-infrared PL relations confirmed the mid-infrared relations' independence from metallicity effects. Additionally, we applied our posterior distance estimates of our 76 RRL sample to fit a new absolute $V$-band luminosity-metallicity relation.

Perhaps the most significant contribution possible of the RRL PL relation is a well-constrained measurement of the LMC distance. The distance modulus of the LMC is a hugely consequential value in the extension of the distance ladder out to cosmological scales, and the the subsequent calculation of the Hubble constant, $H_0$ \citep{2008AJ....135..112S}. The mid-infrared PL relations presented here will allow future studies of LMC RRL variables conducted with {\it Spitzer} (warm) or possibly the James Webb Space Telescope (JWST) to measure reliable LMC distances with error at the $\sim$2\% level or lower\footnote{Note that the current absolute calibration uncertainty of WISE relative to Spitzer is  2.4, 2.8, 4.5\% (W1,W2,W3, respectively), as provided in the Explanatory Supplement to the WISE Preliminary Data Release Products --- \url{http://wise2.ipac.caltech.edu/docs/release/prelim/expsup/sec4_3g.html}. This would dominate over the errors in our WISE-determined distance measure.}. It is conceivable that a comprehensive mid-infrared survey of LMC RRL variables would enable the three-dimensional stellar structure mapping of the LMC with $\sim$1 kpc resolution.

The accuracy of any estimate of the PL relation is influenced by the accuracy of the {\it a priori} distances for the RRL sample used. Soon the HST parallax measurements \citep{2008hst..prop11789B} of V*RZCep, V*UVOct, V*SUDra and V*XZCyg will be published. Our results in Table \ref{distances_table} serve as predictions of what will be found for the first three of those sources (once the WISE data on V*XZCyg is released, eqs.~\ref{eq:ans} could be used to postdict the HST result). The Gaia satellite of the  European Space Agency, a 5-year astrometry mission to be launched in mid-2013, promises trigonometric parallax measurements of all field RRL variables within 3 kpc with individual accuracy $\sigma (\pi)<3\%$ \citep{2009IAUS..258..409C}. Although these measurements will not be available for many years to come, they have tremendous potential to further constrain the PL relations presented herein. In doing so, we can hope to study Galactic substructure well into the optically-obscured Galactic plane and
further improve the resulting distance estimates for the LMC and beyond.

\bigskip

\noindent {\bf Acknowledgements:} We thank D.\ Hoffman, R.\ Cutri,  P.\ Eisenhardt, and N.\ Wright for valuable conversations about WISE and the WISE data. We thank the entire WISE team for having produced a wonderful mid-IR dataset. The authors acknowledge the generous support of a CDI grant (\#0941742) from the National Science Foundation. JSB and CRK were also partially supported by grant NSF/AST-100991.  NRB is supported through the Einstein Fellowship Program (NASA Cooperative Agreement: NNG06DO90A).  This research has made use of the NASA/IPAC Infrared Science Archive, which is operated by the Jet Propulsion Laboratory, California Institute of Technology, under contract with the National Aeronautics and Space Administration. This publication makes use of data products from the Wide-field Infrared Survey Explorer, which is a joint project of the University of California, Los Angeles, and the Jet Propulsion Laboratory/California Institute of Technology, funded by the National Aeronautics and Space Administration.

\appendix

\section{WISE Period Recovery}\label{p_rec}

\begin{figure}[p]
\centerline{\includegraphics[width=4in]{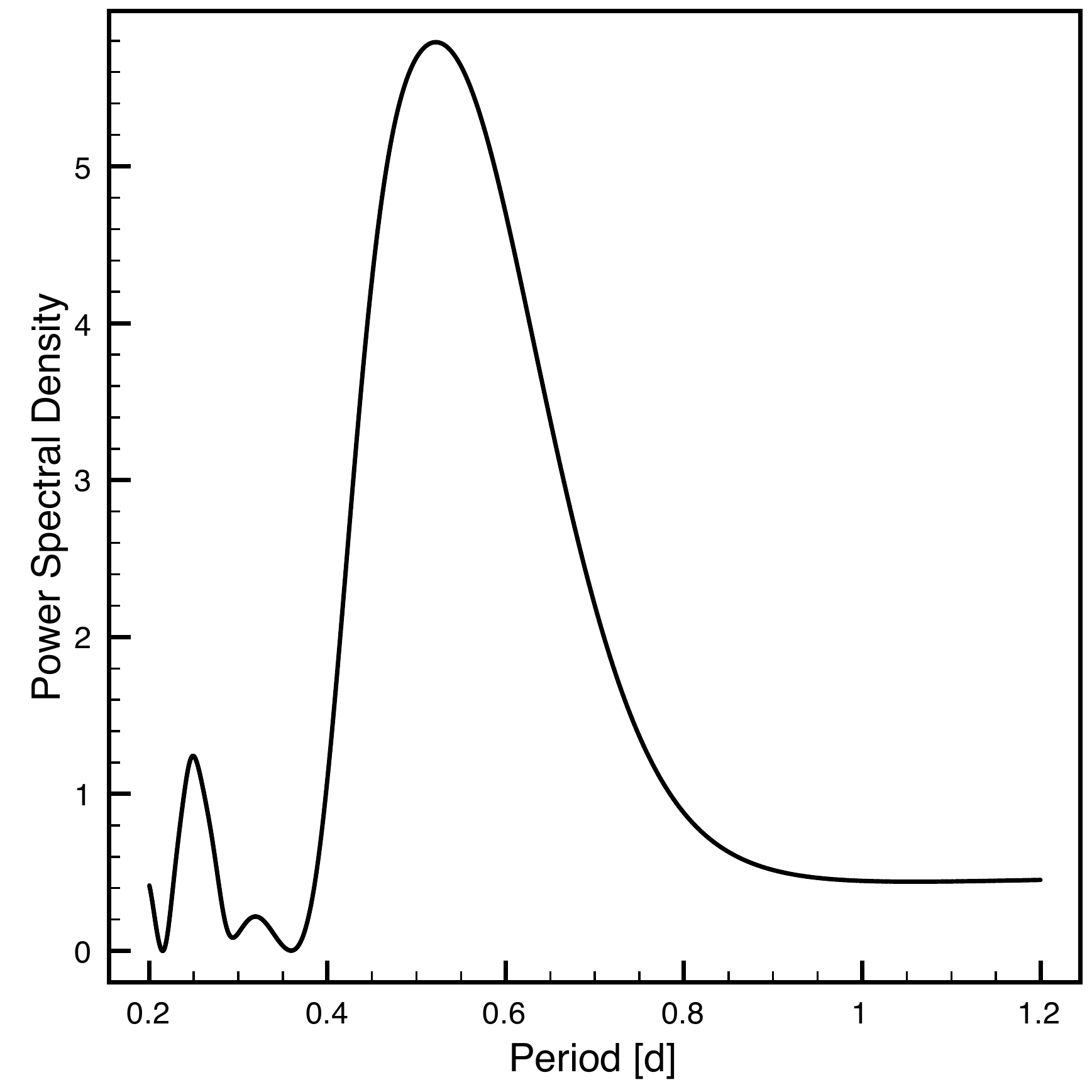}}
\caption{Periodogram generated from the 14 epochs of W2 data for V*MSAra. To envelope the reasonable period range of RRL we plot from 0.2 to 1.2 day. The archival \hipp\ period of 0.525 day is well recovered from the WISE data (peak at 0.522 day).}
\label{periodogram}
\end{figure}

There are two primary concerns with using WISE data to discover short-period variable stars. First, the number of observations on any given patch of sky is small (minimum 16 in the final WISE dataset) and determined primarily by the ecliptic latitude. Secondly, the peak-to-trough amplitude of pulsating variables is significantly decreased at mid-infrared wavelengths as compared to optical wavelengths ($\sim$0.2 mag in W1 compared to $\sim$1 mag in $V$ for RRL variables). Our analysis shows that even with these disadvantageous factors, the WISE light curves can yield accurate periods quite often. Peaks in the periodogram are
expected to have frequency widths $\sim 1/T$, where $T$ is the time spanned
by the observations.  We note that our best-fit frequencies, determined
on a grid of frequency steps $0.01/T$,
agree with well with those of \cite{1998AA...330..515F} 
(to better than $0.2/T$ typically). We plot an example periodogram using the W2 light curve of an RRL with the median number of WISE observations (14) in Figure \ref{periodogram}.

For the fitting of PL relations, it is important to have accurate log-Period estimates or, equivalently, accurate fractional period estimates. We describe the (in)accuracy of a recovered period by the simple fractional error as compared to the known, true period.
\beq \label{p_error_eqn}
{\rm Recovered~Period~Fract.~Error} = \left|{\rm P_{m}} - {\rm P_{t}}\right| /{\rm P_{t}}
\eeq
with ${\rm P_{m}}$ the period measured solely from the WISE light curve and ${\rm P_{t}}$ the true period as measured from the \hipp\ light curve.

\begin{figure}[p]
\centerline{\includegraphics[width=4in]{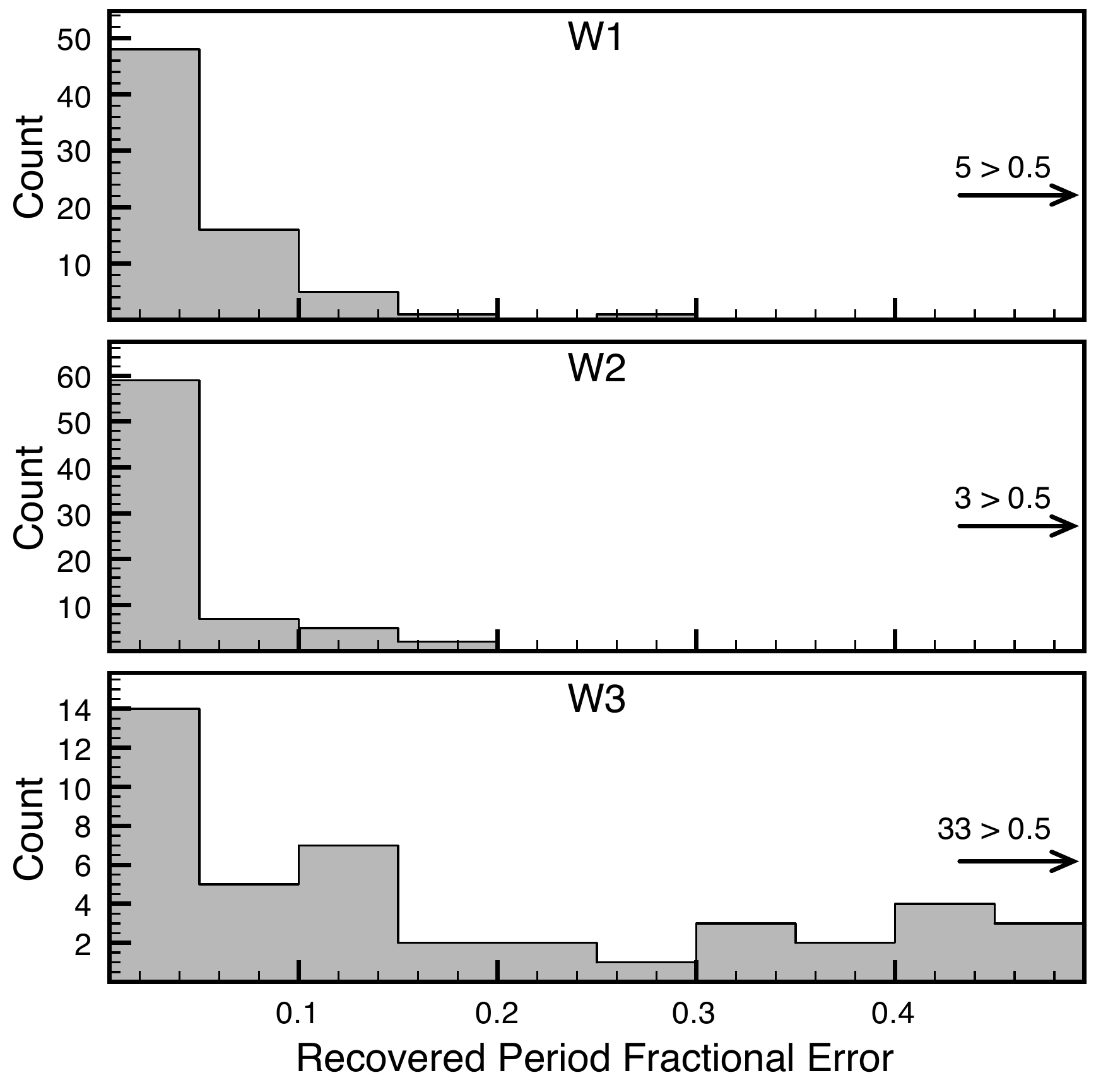}}
\caption{Distribution of recovered period fractional error across the three WISE bands. The plots are cropped to a maximum fractional error of 0.5 and the resultant number of excluded light curves is noted in each plot.}
\label{period_recovery_histogram}
\end{figure}

Figure \ref{period_recovery_histogram} shows histograms of the period recovery accuracy for each WISE band relative to \hipp\ and illustrates that nearly all WISE light curves produce accurate periods in the shorter wavelength bands W1 and W2.

\begin{figure}[p]
\centerline{\includegraphics[width=4in]{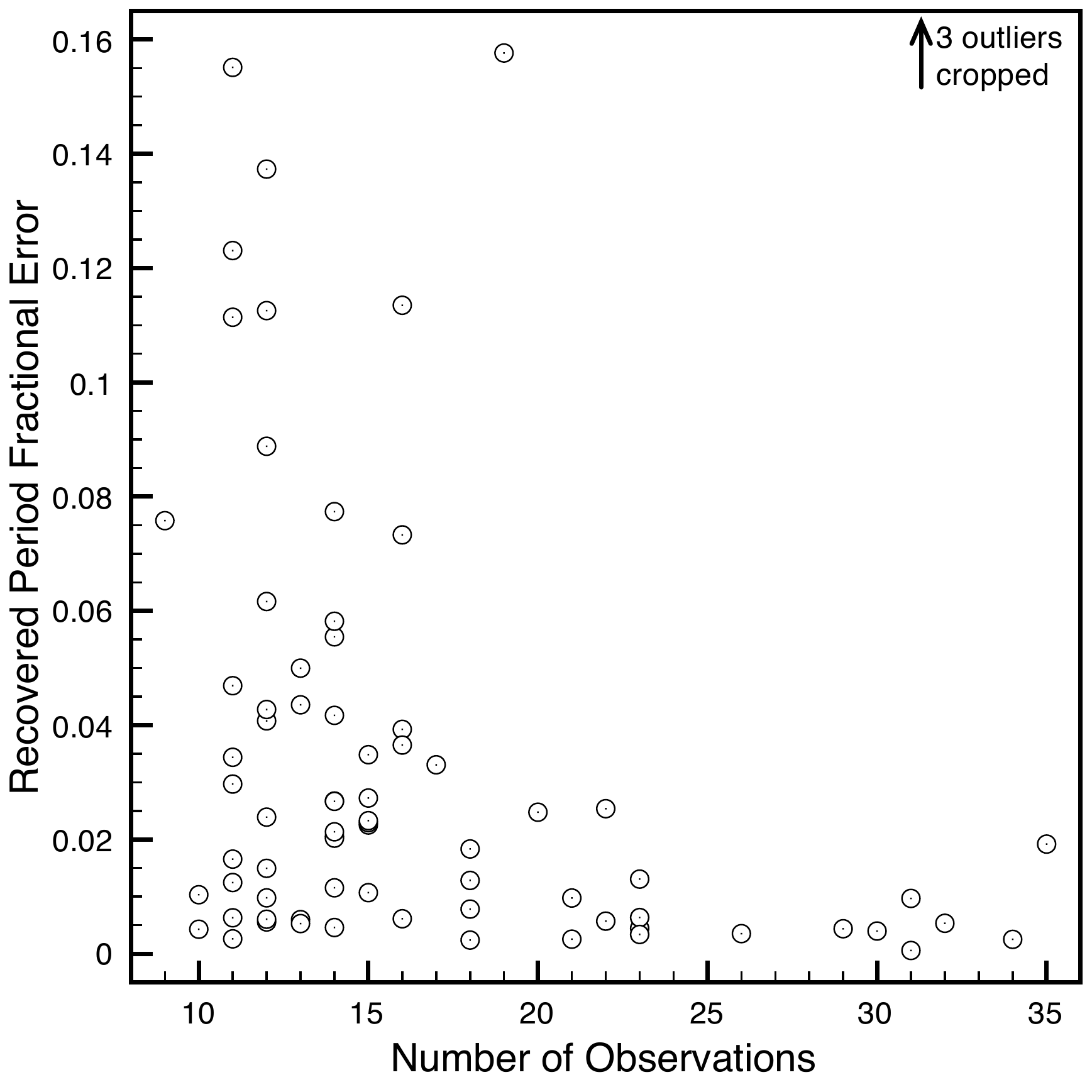}}
\caption{Recovered period fractional error plotted as a function of number of WISE observations for band W2. As expected, there is a general trend of reduced recovered period fractional error with increasing number of observations. Beyond 20 observations, the typical period error is $\lesssim 2$\%.}
\label{fperiod_recovery_numobs}
\end{figure}

To explore how the number of epochs in a light curve affects period recovery, we plot in Figure \ref{fperiod_recovery_numobs} recovered period fractional error as a function of the number of observations for band W2. Because of the survey strategy of WISE, a larger number of observations typically indicates both increased temporal resolution (increased frequency of observation) and increased total light curve timespan (duration between first and last observation). As expected, there is a general trend of reduced recovered period fractional error with increasing number of observations. Beyond 20 observations, the typical period error is $\lesssim 2$\%.

\begin{figure}[p]
\centerline{\includegraphics[width=4in]{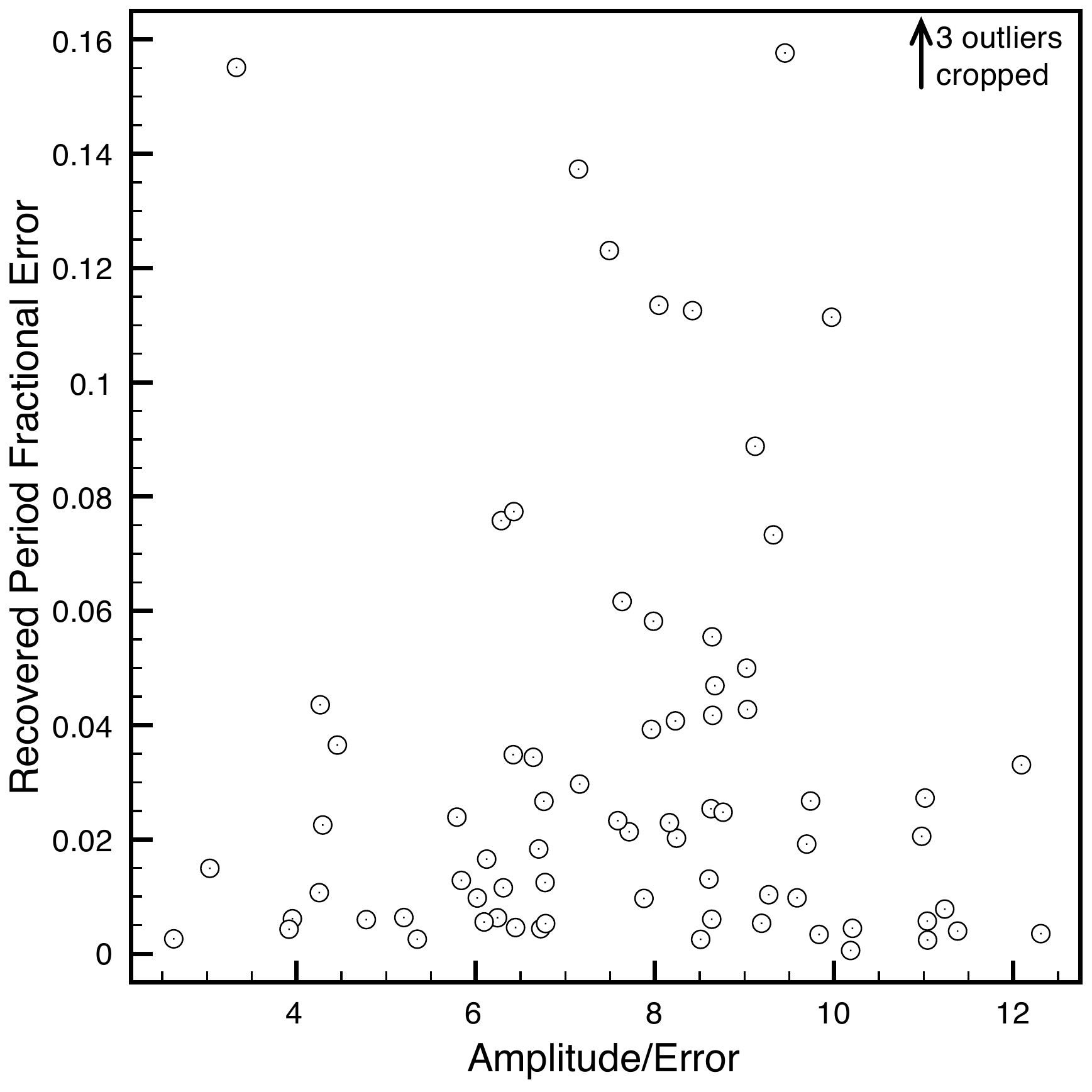}}
\caption{Recovered period fractional error plotted as a function of light curve ${\rm amplitude}/\langle{\rm mag~error}\rangle$ for band W2. Although we would expect reduced recovered period fractional error with increasing ${\rm amplitude}/\langle{\rm mag~error}\rangle$, this trend is not obvious in the plot. ``Vertical light curve smudging'' is not a significant source of error in the recovered period for our dataset.}
\label{period_recovery_aovererror}
\end{figure}

Any period-finding algorithm must distinguish a shape for the light curve. That is, a phased light curve must be smoothly varying in that the uncertainty in the brightness at any phase point is considerably smaller than the amplitude of the light curve. As the photometric uncertainty increases relative to the amplitude, there is an effect of ``vertical smudging'' in which the light curve shape becomes less distinguishable. The flux amplitudes of RRL variables are about two times smaller in the mid-infrared as compared to the visual band. To investigate if this plays a factor in period recovery with WISE light curve data, Figure \ref{period_recovery_aovererror} plots recovered period fractional error as a function of light curve ${\rm amplitude}/\langle{\rm mag~error}\rangle$ for band W2. Although we would expect to observe decreased period error with increased ${\rm amplitude}/\langle{\rm mag~error}\rangle$, this is not observed. We can conclude that ``vertical smudging'' is at most a non-dominant source of error in the recovered periods.

\clearpage

\bibliography{Klein_refs}

\clearpage

\def\fern{a}
\def\here{b}
\def\av{c}
\def\mv{d}
\def\wise{e}
\def\rrlmu{f}
\def\blaz{g}

\begin{deluxetable}{llrrc@{$\pm$}lc@{$\pm$}lr@{$\pm$}lr@{$\pm$}lr@{$\pm$}lr@{$\pm$}lr@{$\pm$}l}
\rotate
\tabletypesize{\scriptsize}
\tablewidth{8.20in}
\tablecolumns{18}
\tablecaption{Input into the PL relation fit. \label{inputs_table}}
\tablehead{   
 \colhead{Name} &
 \colhead{Class\tablenotemark{\fern}} &
 \colhead{[Fe/H]\tablenotemark{\fern}} &
 \colhead{Period\tablenotemark{\here}} &
 \multicolumn{2}{c}{$A_{V,{\rm eff}}$\tablenotemark{\av}} &
 \multicolumn{2}{c}{${m}^*_{V}$\tablenotemark{\mv}} &
 \multicolumn{2}{c}{$\mu$} &
 \multicolumn{2}{c}{${M}_{V}$} &
 \multicolumn{2}{c}{${m}_{\rm W1}$\tablenotemark{\wise}} &
 \multicolumn{2}{c}{${m}_{\rm W2}$\tablenotemark{\wise}} &
 \multicolumn{2}{c}{${m}_{\rm W3}$\tablenotemark{\wise}} \\
\colhead{} &
\colhead{} &
\colhead{} &
\colhead{[d]} & 
\multicolumn{2}{c}{[mag]} &
\multicolumn{2}{c}{[mag]} &
\multicolumn{2}{c}{[mag]} &
\multicolumn{2}{c}{[mag]} &
\multicolumn{2}{c}{[mag]} &
\multicolumn{2}{c}{[mag]} &
\multicolumn{2}{c}{[mag]} }
\startdata
V*AACMi & RRab & $-0.150$ & $0.476323$ & $0.237$ & $0.007$ & $11.337$ & $0.011$ & $10.443$ & $0.138$ & $0.893$ & $0.138$ & $10.240$ & $0.008$ & $10.232$ & $0.049$ & $10.171$ & $0.008$\\
V*AEBoo & RRc & $-1.390$ & $0.314896$ & $0.083$ & $0.004$ & $10.577$ & $0.008$ & $9.969$ & $0.125$ & $0.608$ & $0.125$ & $9.722$ & $0.008$ & $9.730$ & $0.033$ & $9.665$ & $0.008$\\
V*AFVir & RRab & $-1.330$ & $0.483722$ & $0.075$ & $0.004$ & $11.726$ & $0.012$ & $11.104$ & $0.126$ & $0.622$ & $0.125$ & $10.683$ & $0.009$ & $10.689$ & $0.074$ & $10.430$ & $0.008$\\
V*AMVir\tablenotemark{\blaz} & RRab & $-1.370$ & $0.615088$ & $0.226$ & $0.005$ & $11.293$ & $0.010$ & $10.680$ & $0.126$ & $0.613$ & $0.125$ & $10.101$ & $0.009$ & $10.089$ & $0.058$ & $10.030$ & $0.009$\\
V*ANSer & RRab & $-0.070$ & $0.522069$ & $0.130$ & $0.002$ & $10.816$ & $0.010$ & $9.904$ & $0.139$ & $0.912$ & $0.139$ & $9.787$ & $0.008$ & $9.802$ & $0.033$ & $9.759$ & $0.008$\\
V*APSer & RRc & $-1.580$ & $0.340789$ & $0.130$ & $0.003$ & $10.981$ & $0.007$ & $10.417$ & $0.125$ & $0.565$ & $0.125$ & $10.171$ & $0.007$ & $10.177$ & $0.047$ & $10.118$ & $0.007$\\
V*ARHer\tablenotemark{\blaz} & RRab & $-1.300$ & $0.469970$ & $0.038$ & $0.001$ & $11.247$ & $0.008$ & $10.618$ & $0.126$ & $0.629$ & $0.125$ & $10.271$ & $0.007$ & $10.264$ & $0.042$ & $10.234$ & $0.007$\\
V*ARPer & RRab & $-0.300$ & $0.425549$ & $0.571$ & $0.012$ & $9.909$ & $0.014$ & $9.050$ & $0.136$ & $0.859$ & $0.135$ & $8.568$ & $0.007$ & $8.574$ & $0.014$ & $8.534$ & $0.006$\\
V*ATSer & RRab & $-2.030$ & $0.746568$ & $0.123$ & $0.002$ & $11.356$ & $0.011$ & $10.895$ & $0.126$ & $0.461$ & $0.126$ & $10.158$ & $0.007$ & $10.164$ & $0.049$ & $10.196$ & $0.007$\\
V*AUVir & RRc & $-1.500$ & $0.343230$ & $0.091$ & $0.001$ & $11.514$ & $0.009$ & $10.931$ & $0.125$ & $0.583$ & $0.125$ & $10.758$ & $0.009$ & $10.764$ & $0.106$ & $10.612$ & $0.010$\\
V*BBEri & RRab & $-1.320$ & $0.569896$ & $0.154$ & $0.006$ & $11.363$ & $0.010$ & $10.739$ & $0.126$ & $0.624$ & $0.125$ & $10.200$ & $0.005$ & $10.199$ & $0.030$ & $10.099$ & $0.005$\\
V*BCDra & RRab & $-2.000$ & $0.719576$ & $0.210$ & $0.011$ & $11.377$ & $0.013$ & $10.909$ & $0.127$ & $0.468$ & $0.126$ & $10.082$ & $0.005$ & $10.082$ & $0.027$ & $9.967$ & $0.005$\\
V*BNPav & RRab & $-1.320$ & $0.567173$ & $0.260$ & $0.007$ & $12.304$ & $0.018$ & $11.679$ & $0.127$ & $0.624$ & $0.125$ & $11.282$ & $0.014$ & $11.282$ & $0.105$ & $11.079$ & $0.012$\\
V*BNVul & RRab & $-1.610$ & $0.594125$ & $0.983$ & $0.033$ & $9.983$ & $0.034$ & $9.425$ & $0.130$ & $0.558$ & $0.125$ & $8.635$ & $0.007$ & $8.605$ & $0.013$ & $8.529$ & $0.006$\\
V*BPPav & RRab & $-1.480$ & $0.527128$ & $0.202$ & $0.008$ & $12.347$ & $0.019$ & $11.759$ & $0.126$ & $0.588$ & $0.125$ & $11.303$ & $0.006$ & $11.300$ & $0.082$ & $11.073$ & $0.006$\\
V*CGLib & RRc & $-1.190$ & $0.306789$ & $0.686$ & $0.045$ & $10.828$ & $0.046$ & $10.173$ & $0.134$ & $0.654$ & $0.126$ & $10.132$ & $0.008$ & $10.128$ & $0.055$ & $10.063$ & $0.008$\\
V*CIAnd & RRab & $-0.690$ & $0.484718$ & $0.208$ & $0.007$ & $12.077$ & $0.013$ & $11.308$ & $0.131$ & $0.769$ & $0.130$ & $11.000$ & $0.009$ & $10.994$ & $0.095$ & $10.896$ & $0.009$\\
V*CNLyr & RRab & $-0.580$ & $0.411382$ & $0.544$ & $0.023$ & $10.900$ & $0.024$ & $10.106$ & $0.133$ & $0.795$ & $0.131$ & $9.856$ & $0.006$ & $9.861$ & $0.031$ & $9.816$ & $0.007$\\
V*DDHya & RRab & $-0.970$ & $0.501818$ & $0.075$ & $0.002$ & $12.169$ & $0.016$ & $11.464$ & $0.128$ & $0.705$ & $0.127$ & $11.105$ & $0.009$ & $11.089$ & $0.113$ & $10.777$ & $0.009$\\
V*FWLup & RRab & $-0.200$ & $0.484171$ & $0.280$ & $0.009$ & $8.758$ & $0.011$ & $7.876$ & $0.137$ & $0.882$ & $0.137$ & $7.645$ & $0.007$ & $7.661$ & $0.009$ & $7.627$ & $0.007$\\
V*HHPup & RRab & $-0.500$ & $0.390746$ & $0.474$ & $0.017$ & $10.794$ & $0.019$ & $9.981$ & $0.134$ & $0.813$ & $0.132$ & $9.887$ & $0.006$ & $9.881$ & $0.021$ & $9.749$ & $0.006$\\
V*HKPup & RRab & $-1.110$ & $0.734254$ & $0.576$ & $0.021$ & $10.753$ & $0.022$ & $10.080$ & $0.128$ & $0.673$ & $0.126$ & $9.866$ & $0.008$ & $9.854$ & $0.032$ & $9.769$ & $0.007$\\
V*IOLyr & RRab & $-1.140$ & $0.577122$ & $0.206$ & $0.014$ & $11.641$ & $0.016$ & $10.975$ & $0.127$ & $0.666$ & $0.126$ & $10.501$ & $0.006$ & $10.496$ & $0.043$ & $10.393$ & $0.006$\\
V*MSAra & RRab & $-1.480$ & $0.524958$ & $0.369$ & $0.006$ & $11.688$ & $0.012$ & $11.101$ & $0.126$ & $0.588$ & $0.125$ & $10.598$ & $0.008$ & $10.588$ & $0.068$ & $10.511$ & $0.008$\\
V*MTTel & RRc & $-1.850$ & $0.316899$ & $0.138$ & $0.007$ & $8.844$ & $0.010$ & $8.341$ & $0.126$ & $0.502$ & $0.125$ & $8.067$ & $0.007$ & $8.077$ & $0.011$ & $8.036$ & $0.007$\\
V*RRGem\tablenotemark{\blaz} & RRab & $-0.290$ & $0.397316$ & $0.201$ & $0.003$ & $11.159$ & $0.012$ & $10.297$ & $0.136$ & $0.861$ & $0.135$ & $10.226$ & $0.008$ & $10.223$ & $0.052$ & $10.046$ & $0.008$\\
V*RRLyr\tablenotemark{\blaz} & RRab & $-1.390$ & $0.566805$ & $0.102$ & $0.003$ & $7.650$ & $0.007$ & $7.090$ & $0.063$\tablenotemark{\rrlmu} & $0.608$ & $0.125$ & $6.519$ & $0.008$ & $6.486$ & $0.006$ & $6.431$ & $0.005$\\
V*RSBoo\tablenotemark{\blaz} & RRab & $-0.360$ & $0.377337$ & $0.044$ & $0.003$ & $10.339$ & $0.007$ & $9.494$ & $0.135$ & $0.845$ & $0.134$ & $9.411$ & $0.009$ & $9.398$ & $0.028$ & $9.386$ & $0.009$\\
V*RVCet\tablenotemark{\blaz} & RRab & $-1.600$ & $0.623428$ & $0.097$ & $0.002$ & $10.828$ & $0.007$ & $10.268$ & $0.120$ & $0.560$ & $0.120$ & $9.568$ & $0.012$ & $9.563$ & $0.044$ & $9.483$ & $0.011$\\
V*RVCrB & RRc & $-1.690$ & $0.331593$ & $0.134$ & $0.005$ & $11.284$ & $0.008$ & $10.744$ & $0.125$ & $0.539$ & $0.125$ & $10.468$ & $0.007$ & $10.481$ & $0.051$ & $10.365$ & $0.008$\\
V*RVOct & RRab & $-1.710$ & $0.571130$ & $0.520$ & $0.035$ & $10.434$ & $0.036$ & $9.899$ & $0.130$ & $0.535$ & $0.125$ & $9.449$ & $0.006$ & $9.444$ & $0.020$ & $9.385$ & $0.006$\\
V*RWCnc\tablenotemark{\blaz} & RRab & $-1.670$ & $0.547193$ & $0.067$ & $0.003$ & $11.788$ & $0.017$ & $11.244$ & $0.126$ & $0.544$ & $0.125$ & $10.665$ & $0.008$ & $10.683$ & $0.066$ & $10.456$ & $0.008$\\
V*RWDra\tablenotemark{\blaz} & RRab & $-1.550$ & $0.442898$ & $0.040$ & $0.002$ & $11.709$ & $0.009$ & $11.137$ & $0.125$ & $0.572$ & $0.125$ & $10.662$ & $0.005$ & $10.664$ & $0.039$ & $10.545$ & $0.005$\\
V*RWTrA & RRab & $-0.130$ & $0.374035$ & $0.258$ & $0.007$ & $11.129$ & $0.010$ & $10.231$ & $0.138$ & $0.898$ & $0.138$ & $10.041$ & $0.007$ & $10.038$ & $0.043$ & $10.112$ & $0.007$\\
V*RXCol\tablenotemark{\blaz} & RRab & $-1.700$ & $0.593733$ & $0.243$ & $0.006$ & $12.420$ & $0.022$ & $11.883$ & $0.127$ & $0.537$ & $0.125$ & $11.241$ & $0.006$ & $11.232$ & $0.098$ & $10.964$ & $0.007$\\
V*RXEri & RRab & $-1.330$ & $0.587246$ & $0.194$ & $0.006$ & $9.484$ & $0.008$ & $8.862$ & $0.126$ & $0.622$ & $0.125$ & $8.316$ & $0.005$ & $8.318$ & $0.008$ & $8.243$ & $0.005$\\
V*RYCol\tablenotemark{\blaz} & RRab & $-0.910$ & $0.478857$ & $0.090$ & $0.004$ & $10.830$ & $0.008$ & $10.111$ & $0.128$ & $0.719$ & $0.128$ & $9.786$ & $0.005$ & $9.794$ & $0.021$ & $9.728$ & $0.005$\\
V*RYOct & RRab & $-1.830$ & $0.563469$ & $0.354$ & $0.010$ & $11.675$ & $0.013$ & $11.168$ & $0.126$ & $0.507$ & $0.125$ & $10.725$ & $0.007$ & $10.718$ & $0.063$ & $10.628$ & $0.007$\\
V*RZCep & RRc & $-1.770$ & $0.308645$ & $0.864$ & $0.014$ & $8.557$ & $0.015$ & $8.036$ & $0.126$ & $0.521$ & $0.125$ & $7.858$ & $0.006$ & $7.871$ & $0.009$ & $7.710$ & $0.006$\\
V*RZCet & RRab & $-1.360$ & $0.510613$ & $0.096$ & $0.003$ & $11.721$ & $0.012$ & $11.106$ & $0.126$ & $0.615$ & $0.125$ & $10.587$ & $0.009$ & $10.595$ & $0.071$ & $10.456$ & $0.009$\\
V*SAra\tablenotemark{\blaz} & RRab & $-0.710$ & $0.451888$ & $0.347$ & $0.007$ & $10.423$ & $0.010$ & $9.658$ & $0.130$ & $0.765$ & $0.130$ & $9.521$ & $0.008$ & $9.526$ & $0.029$ & $9.444$ & $0.008$\\
V*SSOct\tablenotemark{\blaz} & RRab & $-1.600$ & $0.621825$ & $1.008$ & $0.020$ & $10.862$ & $0.021$ & $10.302$ & $0.122$ & $0.560$ & $0.120$ & $9.729$ & $0.006$ & $9.704$ & $0.024$ & $9.610$ & $0.006$\\
V*STBoo\tablenotemark{\blaz} & RRab & $-1.760$ & $0.622291$ & $0.062$ & $0.001$ & $10.947$ & $0.007$ & $10.424$ & $0.125$ & $0.523$ & $0.125$ & $9.798$ & $0.010$ & $9.787$ & $0.035$ & $9.740$ & $0.009$\\
V*STVir & RRab & $-0.670$ & $0.410820$ & $0.129$ & $0.002$ & $11.397$ & $0.011$ & $10.624$ & $0.131$ & $0.774$ & $0.130$ & $10.562$ & $0.009$ & $10.556$ & $0.079$ & $10.502$ & $0.008$\\
V*SUDra & RRab & $-1.800$ & $0.660419$ & $0.030$ & $0.002$ & $9.726$ & $0.008$ & $9.212$ & $0.125$ & $0.514$ & $0.125$ & $8.593$ & $0.006$ & $8.585$ & $0.015$ & $8.514$ & $0.007$\\
V*SVEri & RRab & $-1.700$ & $0.713863$ & $0.282$ & $0.008$ & $9.653$ & $0.012$ & $9.116$ & $0.126$ & $0.537$ & $0.125$ & $8.546$ & $0.008$ & $8.553$ & $0.015$ & $8.483$ & $0.007$\\
V*SXFor & RRab & $-1.660$ & $0.605333$ & $0.044$ & $0.003$ & $11.075$ & $0.007$ & $10.528$ & $0.125$ & $0.546$ & $0.125$ & $9.829$ & $0.007$ & $9.832$ & $0.031$ & $9.744$ & $0.007$\\
V*SZGem & RRab & $-1.460$ & $0.501136$ & $0.135$ & $0.002$ & $11.609$ & $0.012$ & $11.016$ & $0.126$ & $0.592$ & $0.125$ & $10.637$ & $0.009$ & $10.623$ & $0.077$ & $10.470$ & $0.009$\\
V*TTCnc\tablenotemark{\blaz} & RRab & $-1.570$ & $0.563450$ & $0.199$ & $0.006$ & $11.122$ & $0.013$ & $10.555$ & $0.126$ & $0.567$ & $0.125$ & $9.935$ & $0.007$ & $9.930$ & $0.043$ & $9.952$ & $0.006$\\
V*TTLyn & RRab & $-1.560$ & $0.597438$ & $0.055$ & $0.002$ & $9.809$ & $0.010$ & $9.240$ & $0.125$ & $0.569$ & $0.125$ & $8.568$ & $0.008$ & $8.573$ & $0.014$ & $8.502$ & $0.007$\\
V*TVCrB & RRab & $-2.330$ & $0.584611$ & $0.122$ & $0.004$ & $11.755$ & $0.011$ & $11.363$ & $0.129$ & $0.392$ & $0.128$ & $10.704$ & $0.008$ & $10.690$ & $0.070$ & $10.663$ & $0.008$\\
V*TWHer & RRab & $-0.690$ & $0.399601$ & $0.131$ & $0.004$ & $11.143$ & $0.008$ & $10.374$ & $0.130$ & $0.769$ & $0.130$ & $10.231$ & $0.006$ & $10.220$ & $0.037$ & $10.114$ & $0.006$\\
V*TWLyn & RRab & $-0.660$ & $0.481853$ & $0.143$ & $0.002$ & $11.855$ & $0.015$ & $11.079$ & $0.131$ & $0.776$ & $0.130$ & $10.727$ & $0.008$ & $10.735$ & $0.088$ & $10.629$ & $0.009$\\
V*TYAps & RRab & $-0.950$ & $0.501692$ & $0.414$ & $0.019$ & $11.429$ & $0.020$ & $10.719$ & $0.129$ & $0.710$ & $0.128$ & $10.346$ & $0.006$ & $10.338$ & $0.042$ & $10.278$ & $0.006$\\
V*TZAur & RRab & $-0.790$ & $0.391675$ & $0.197$ & $0.003$ & $11.717$ & $0.014$ & $10.970$ & $0.130$ & $0.746$ & $0.129$ & $10.780$ & $0.008$ & $10.795$ & $0.092$ & $10.707$ & $0.008$\\
V*ULep & RRab & $-1.780$ & $0.581474$ & $0.103$ & $0.005$ & $10.465$ & $0.009$ & $9.947$ & $0.125$ & $0.519$ & $0.125$ & $9.405$ & $0.005$ & $9.415$ & $0.020$ & $9.405$ & $0.005$\\
V*UPic & RRab & $-0.720$ & $0.440371$ & $0.031$ & $0.001$ & $11.367$ & $0.010$ & $10.604$ & $0.130$ & $0.762$ & $0.130$ & $10.332$ & $0.005$ & $10.337$ & $0.032$ & $10.317$ & $0.005$\\
V*UVOct\tablenotemark{\blaz} & RRab & $-1.740$ & $0.542625$ & $0.250$ & $0.002$ & $9.243$ & $0.006$ & $8.715$ & $0.125$ & $0.528$ & $0.125$ & $8.175$ & $0.006$ & $8.167$ & $0.008$ & $8.077$ & $0.005$\\
V*UYBoo & RRab & $-2.560$ & $0.650845$ & $0.113$ & $0.004$ & $10.854$ & $0.007$ & $10.514$ & $0.131$ & $0.339$ & $0.131$ & $9.706$ & $0.009$ & $9.691$ & $0.034$ & $9.592$ & $0.009$\\
V*UYCam & RRc & $-1.330$ & $0.267042$ & $0.079$ & $0.003$ & $11.461$ & $0.007$ & $10.838$ & $0.126$ & $0.622$ & $0.125$ & $10.822$ & $0.009$ & $10.847$ & $0.092$ & $10.991$ & $0.008$\\
V*V413CrA & RRab & $-1.260$ & $0.589324$ & $0.258$ & $0.010$ & $10.333$ & $0.012$ & $9.695$ & $0.126$ & $0.638$ & $0.126$ & $9.111$ & $0.008$ & $9.107$ & $0.023$ & $9.059$ & $0.007$\\
V*V440Sgr & RRab & $-1.400$ & $0.477474$ & $0.263$ & $0.006$ & $10.051$ & $0.012$ & $9.445$ & $0.126$ & $0.606$ & $0.125$ & $9.040$ & $0.009$ & $9.054$ & $0.027$ & $8.981$ & $0.008$\\
V*V445Oph & RRab & $-0.190$ & $0.397023$ & $0.928$ & $0.027$ & $10.082$ & $0.030$ & $9.198$ & $0.140$ & $0.884$ & $0.137$ & $9.191$ & $0.008$ & $9.183$ & $0.023$ & $9.168$ & $0.007$\\
V*V455Oph & RRab & $-1.070$ & $0.453918$ & $0.454$ & $0.016$ & $11.877$ & $0.019$ & $11.195$ & $0.128$ & $0.682$ & $0.127$ & $10.947$ & $0.008$ & $10.949$ & $0.094$ & $10.851$ & $0.008$\\
V*V499Cen & RRab & $-1.430$ & $0.521210$ & $0.241$ & $0.007$ & $10.863$ & $0.012$ & $10.263$ & $0.126$ & $0.599$ & $0.125$ & $9.830$ & $0.013$ & $9.815$ & $0.036$ & $9.744$ & $0.009$\\
V*V675Sgr & RRab & $-2.280$ & $0.642289$ & $0.424$ & $0.011$ & $9.877$ & $0.014$ & $9.473$ & $0.129$ & $0.404$ & $0.128$ & $8.912$ & $0.008$ & $8.915$ & $0.022$ & $8.892$ & $0.008$\\
V*VInd & RRab & $-1.500$ & $0.479591$ & $0.148$ & $0.003$ & $9.809$ & $0.008$ & $9.226$ & $0.125$ & $0.583$ & $0.125$ & $8.827$ & $0.009$ & $8.821$ & $0.016$ & $8.792$ & $0.009$\\
V*VXHer & RRab & $-1.580$ & $0.455373$ & $0.151$ & $0.006$ & $10.532$ & $0.009$ & $9.967$ & $0.125$ & $0.565$ & $0.125$ & $9.589$ & $0.009$ & $9.586$ & $0.027$ & $9.478$ & $0.007$\\
V*VYLib & RRab & $-1.340$ & $0.533938$ & $0.561$ & $0.018$ & $11.137$ & $0.020$ & $10.518$ & $0.127$ & $0.620$ & $0.125$ & $10.044$ & $0.009$ & $10.001$ & $0.056$ & $10.008$ & $0.011$\\
V*VYSer & RRab & $-1.790$ & $0.714094$ & $0.129$ & $0.002$ & $9.998$ & $0.008$ & $9.482$ & $0.125$ & $0.516$ & $0.125$ & $8.752$ & $0.008$ & $8.738$ & $0.017$ & $8.678$ & $0.007$\\
V*VZHer & RRab & $-1.020$ & $0.440326$ & $0.096$ & $0.002$ & $11.389$ & $0.008$ & $10.696$ & $0.127$ & $0.693$ & $0.127$ & $10.475$ & $0.007$ & $10.463$ & $0.048$ & $10.397$ & $0.006$\\
V*WYPav & RRab & $-0.980$ & $0.588580$ & $0.326$ & $0.003$ & $11.854$ & $0.011$ & $11.151$ & $0.128$ & $0.703$ & $0.127$ & $10.542$ & $0.008$ & $10.540$ & $0.051$ & $10.256$ & $0.007$\\
V*XAri & RRab & $-2.430$ & $0.651139$ & $0.610$ & $0.018$ & $8.935$ & $0.022$ & $8.566$ & $0.131$ & $0.369$ & $0.129$ & $7.875$ & $0.007$ & $7.885$ & $0.011$ & $7.815$ & $0.007$\\
V*XXAnd & RRab & $-1.940$ & $0.722742$ & $0.138$ & $0.002$ & $10.541$ & $0.008$ & $10.060$ & $0.126$ & $0.482$ & $0.126$ & $9.374$ & $0.008$ & $9.385$ & $0.025$ & $9.252$ & $0.007$\\
V*XXPup & RRab & $-1.330$ & $0.517198$ & $0.196$ & $0.003$ & $11.042$ & $0.010$ & $10.420$ & $0.126$ & $0.622$ & $0.125$ & $9.996$ & $0.008$ & $9.989$ & $0.038$ & $9.988$ & $0.009$\\
V*XZAps & RRab & $-1.060$ & $0.587277$ & $0.446$ & $0.022$ & $11.937$ & $0.025$ & $11.253$ & $0.129$ & $0.684$ & $0.127$ & $10.835$ & $0.006$ & $10.828$ & $0.067$ & $10.884$ & $0.006$\\
\enddata
\tablenotetext{\fern}{From \citet{1998AA...330..515F}.}
\tablenotetext{\here}{Period determined herein using \hipp\ data.}
\tablenotetext{\av}{Effective extinction using SFD dust models and the \citet{1998ApJ...508..844G} Galactic dust model. See \S \ref{data_desc}.}
\tablenotetext{\mv}{Extinction-corrected apparent magnitude.}
\tablenotetext{\wise}{Determined from the WISE data following \S \ref{meth}.}
\tablenotetext{\rrlmu}{The value of $\mu$ used as a prior in the analysis is from HST parallax measurements \citep{2002AJ....123..473B}. The \hipp-based distance modulus determination (\S \ref{data_desc}) yields $\mu = 7.042 \pm 0.125$.}
\tablenotetext{\blaz}{Blazhko-affected star following from \url{http://www.univie.ac.at/tops/blazhko/Blazhkolist.html}}
\end{deluxetable}

\def\fromfit{a}
\def\ddesc{b}
\def\calcm{c}

\begin{deluxetable}{lccccr@{$\pm$}l}
\tabletypesize{\scriptsize}
\tablewidth{4.90in}
\setlength{\tabcolsep}{0.07in}
\tablecolumns{7}
\tablecaption{RRL Distance and $M_V$ Posteriors from Bayesian Analysis\label{distances_table}}
\tablehead{   
 \colhead{Name} &
 \colhead{${\rm d}_{\rm best}$ \tablenotemark{\fromfit}} &
 \colhead{$[{\rm d}-1\sigma, {\rm d}+1\sigma]$} &
 \colhead{$[{\rm d}-2\sigma, {\rm d}+2\sigma]$} &
 \colhead{$\Delta(\mu_{\rm prior}-\mu_{\rm post})$ \tablenotemark{\ddesc}} &
 \multicolumn{2}{c}{${M}_{V}$ \tablenotemark{\calcm}} \\
\colhead{} &
\colhead{[pc]} &
\colhead{[pc]} &
\colhead{[pc]} & 
\colhead{[No. of $\sigma$]} &
\multicolumn{2}{c}{[mag]}
}
\startdata
V*AACMi & $1330.7$ & $[1320.4,1341.1]$ & $[1310.1,1351.6]$ & $1.27$ & $0.716$ & $0.020$\\
V*AEBoo & $914.9$ & $[900.6,929.5]$ & $[886.6,944.2]$ & $-1.25$ & $0.770$ & $0.035$\\
V*AFVir & $1643.8$ & $[1631.2,1656.6]$ & $[1618.6,1669.4]$ & $-0.20$ & $0.647$ & $0.021$\\
V*AMVir & $1360.2$ & $[1347.3,1373.2]$ & $[1334.5,1386.4]$ & $-0.09$ & $0.625$ & $0.023$\\
V*ANSer & $1120.8$ & $[1112.3,1129.3]$ & $[1103.8,1137.9]$ & $2.44$ & $0.569$ & $0.019$\\
V*APSer & $1154.9$ & $[1139.2,1170.8]$ & $[1123.6,1187.0]$ & $-0.81$ & $0.668$ & $0.031$\\
V*ARHer & $1344.8$ & $[1334.4,1355.4]$ & $[1324.0,1366.0]$ & $0.20$ & $0.604$ & $0.019$\\
V*ARPer & $597.1$ & $[591.8,602.4]$ & $[586.6,607.7]$ & $-1.24$ & $1.029$ & $0.024$\\
V*ATSer & $1499.6$ & $[1479.3,1520.3]$ & $[1459.2,1541.2]$ & $-0.11$ & $0.476$ & $0.032$\\
V*AUVir & $1515.7$ & $[1495.2,1536.5]$ & $[1475.0,1557.6]$ & $-0.22$ & $0.611$ & $0.031$\\
V*BBEri & $1389.6$ & $[1378.4,1400.9]$ & $[1367.3,1412.4]$ & $-0.19$ & $0.648$ & $0.020$\\
V*BCDra & $1424.1$ & $[1406.2,1442.2]$ & $[1388.5,1460.6]$ & $-1.09$ & $0.610$ & $0.031$\\
V*BNPav & $2283.5$ & $[2263.4,2303.8]$ & $[2243.4,2324.2]$ & $0.89$ & $0.511$ & $0.026$\\
V*BNVul & $680.4$ & $[674.6,686.4]$ & $[668.7,692.4]$ & $-1.99$ & $0.819$ & $0.039$\\
V*BPPav & $2248.3$ & $[2231.5,2265.3]$ & $[2214.8,2282.3]$ & $0.00$ & $0.587$ & $0.025$\\
V*CGLib & $1091.5$ & $[1073.7,1109.6]$ & $[1056.2,1128.0]$ & $0.12$ & $0.638$ & $0.059$\\
V*CIAnd & $1899.9$ & $[1885.2,1914.8]$ & $[1870.6,1929.7]$ & $0.65$ & $0.683$ & $0.021$\\
V*CNLyr & $1065.1$ & $[1055.0,1075.3]$ & $[1044.9,1085.6]$ & $0.23$ & $0.764$ & $0.032$\\
V*DDHya & $2011.6$ & $[1996.2,2027.2]$ & $[1980.9,2042.9]$ & $0.41$ & $0.651$ & $0.023$\\
V*FWLup & $409.7$ & $[406.7,412.7]$ & $[403.7,415.7]$ & $1.34$ & $0.696$ & $0.019$\\
V*HHPup & $1054.3$ & $[1043.3,1065.5]$ & $[1032.5,1076.7]$ & $0.99$ & $0.679$ & $0.030$\\
V*HKPup & $1294.6$ & $[1277.5,1311.9]$ & $[1260.6,1329.4]$ & $3.66$ & $0.192$ & $0.036$\\
V*IOLyr & $1601.8$ & $[1588.4,1615.4]$ & $[1575.1,1629.0]$ & $0.38$ & $0.618$ & $0.024$\\
V*MSAra & $1621.3$ & $[1608.9,1633.8]$ & $[1596.5,1646.4]$ & $-0.41$ & $0.639$ & $0.021$\\
V*MTTel & $429.7$ & $[423.1,436.4]$ & $[416.6,443.2]$ & $-1.35$ & $0.678$ & $0.035$\\
V*RRGem & $1242.5$ & $[1229.8,1255.4]$ & $[1217.2,1268.3]$ & $1.26$ & $0.687$ & $0.025$\\
V*RRLyr & $252.9$ & $[250.9,254.9]$ & $[248.9,256.9]$ & $-1.16$ & $0.636$ & $0.018$\\
V*RSBoo & $840.2$ & $[830.7,849.8]$ & $[821.3,859.6]$ & $0.93$ & $0.718$ & $0.026$\\
V*RVCet & $1070.8$ & $[1060.1,1081.6]$ & $[1049.6,1092.5]$ & $-0.98$ & $0.680$ & $0.023$\\
V*RVCrB & $1313.1$ & $[1294.3,1332.2]$ & $[1275.7,1351.5]$ & $-1.18$ & $0.692$ & $0.032$\\
V*RVOct & $984.3$ & $[976.3,992.4]$ & $[968.4,1000.5]$ & $0.50$ & $0.469$ & $0.040$\\
V*RWCnc & $1704.9$ & $[1691.3,1718.6]$ & $[1677.8,1732.4]$ & $-0.67$ & $0.629$ & $0.024$\\
V*RWDra & $1579.6$ & $[1566.5,1592.9]$ & $[1553.5,1606.2]$ & $-1.14$ & $0.716$ & $0.020$\\
V*RWTrA & $1122.5$ & $[1109.5,1135.6]$ & $[1096.7,1148.8]$ & $0.14$ & $0.879$ & $0.027$\\
V*RXCol & $2272.3$ & $[2252.3,2292.4]$ & $[2232.5,2312.7]$ & $-0.78$ & $0.637$ & $0.029$\\
V*RXEri & $590.1$ & $[585.1,595.1]$ & $[580.2,600.1]$ & $-0.06$ & $0.630$ & $0.020$\\
V*RYCol & $1086.1$ & $[1078.0,1094.3]$ & $[1070.0,1102.5]$ & $0.53$ & $0.651$ & $0.018$\\
V*RYOct & $1761.4$ & $[1747.2,1775.7]$ & $[1733.0,1790.2]$ & $0.48$ & $0.446$ & $0.022$\\
V*RZCep & $381.3$ & $[375.2,387.6]$ & $[369.2,393.9]$ & $-0.99$ & $0.650$ & $0.038$\\
V*RZCet & $1604.0$ & $[1591.8,1616.3]$ & $[1579.6,1628.8]$ & $-0.63$ & $0.695$ & $0.020$\\
V*SAra & $941.0$ & $[933.3,948.8]$ & $[925.6,956.6]$ & $1.60$ & $0.555$ & $0.020$\\
V*SSOct & $1144.8$ & $[1133.9,1155.8]$ & $[1123.1,1167.0]$ & $-0.07$ & $0.569$ & $0.030$\\
V*STBoo & $1188.8$ & $[1177.2,1200.5]$ & $[1165.7,1212.3]$ & $-0.38$ & $0.572$ & $0.022$\\
V*STVir & $1468.1$ & $[1453.9,1482.4]$ & $[1439.8,1496.9]$ & $1.59$ & $0.564$ & $0.024$\\
V*SUDra & $696.2$ & $[688.8,703.7]$ & $[681.4,711.3]$ & $0.01$ & $0.513$ & $0.025$\\
V*SVEri & $702.4$ & $[693.6,711.2]$ & $[685.0,720.2]$ & $0.91$ & $0.420$ & $0.030$\\
V*SXFor & $1197.5$ & $[1186.6,1208.5]$ & $[1175.9,1219.5]$ & $-1.08$ & $0.683$ & $0.021$\\
V*SZGem & $1622.1$ & $[1609.7,1634.6]$ & $[1597.5,1647.1]$ & $0.27$ & $0.558$ & $0.020$\\
V*TTCnc & $1227.2$ & $[1217.3,1237.3]$ & $[1207.5,1247.3]$ & $-0.87$ & $0.677$ & $0.022$\\
V*TTLyn & $667.7$ & $[661.9,673.7]$ & $[656.0,679.7]$ & $-0.92$ & $0.686$ & $0.022$\\
V*TVCrB & $1764.9$ & $[1749.7,1780.2]$ & $[1734.7,1795.6]$ & $-1.00$ & $0.522$ & $0.022$\\
V*TWHer & $1246.4$ & $[1233.8,1259.1]$ & $[1221.4,1271.9]$ & $0.79$ & $0.665$ & $0.023$\\
V*TWLyn & $1677.6$ & $[1664.7,1690.6]$ & $[1651.9,1703.7]$ & $0.34$ & $0.732$ & $0.023$\\
V*TYAps & $1422.3$ & $[1411.8,1432.9]$ & $[1401.3,1443.6]$ & $0.35$ & $0.664$ & $0.026$\\
V*TZAur & $1605.2$ & $[1588.2,1622.4]$ & $[1571.3,1639.8]$ & $0.44$ & $0.689$ & $0.027$\\
V*ULep & $976.8$ & $[968.6,985.1]$ & $[960.4,993.5]$ & $0.02$ & $0.516$ & $0.020$\\
V*UPic & $1357.3$ & $[1346.1,1368.6]$ & $[1334.9,1380.0]$ & $0.45$ & $0.703$ & $0.021$\\
V*UVOct & $535.8$ & $[531.8,539.8]$ & $[527.8,543.9]$ & $-0.56$ & $0.598$ & $0.017$\\
V*UYBoo & $1154.2$ & $[1141.9,1166.6]$ & $[1129.8,1179.1]$ & $-1.53$ & $0.542$ & $0.024$\\
V*UYCam & $1442.5$ & $[1413.8,1471.7]$ & $[1385.6,1501.6]$ & $-0.32$ & $0.665$ & $0.044$\\
V*V413CrA & $852.3$ & $[844.9,859.8]$ & $[837.5,867.3]$ & $-0.33$ & $0.680$ & $0.022$\\
V*V440Sgr & $770.5$ & $[764.6,776.5]$ & $[758.8,782.5]$ & $-0.09$ & $0.617$ & $0.020$\\
V*V445Oph & $773.5$ & $[765.5,781.5]$ & $[757.7,789.6]$ & $1.72$ & $0.640$ & $0.037$\\
V*V455Oph & $1816.7$ & $[1801.8,1831.8]$ & $[1787.0,1847.0]$ & $0.78$ & $0.581$ & $0.026$\\
V*V499Cen & $1133.1$ & $[1124.2,1142.1]$ & $[1115.4,1151.1]$ & $0.06$ & $0.591$ & $0.021$\\
V*V675Sgr & $803.2$ & $[795.1,811.5]$ & $[787.0,819.8]$ & $0.39$ & $0.353$ & $0.026$\\
V*VInd & $698.9$ & $[693.6,704.2]$ & $[688.3,709.6]$ & $-0.03$ & $0.587$ & $0.018$\\
V*VXHer & $970.8$ & $[963.0,978.7]$ & $[955.2,986.7]$ & $-0.25$ & $0.596$ & $0.020$\\
V*VYLib & $1256.2$ & $[1246.2,1266.2]$ & $[1236.3,1276.3]$ & $-0.17$ & $0.642$ & $0.027$\\
V*VYSer & $768.6$ & $[759.1,778.3]$ & $[749.6,788.1]$ & $-0.41$ & $0.569$ & $0.028$\\
V*VZHer & $1442.2$ & $[1430.0,1454.4]$ & $[1417.9,1466.8]$ & $0.77$ & $0.594$ & $0.020$\\
V*WYPav & $1641.2$ & $[1627.0,1655.7]$ & $[1612.8,1670.2]$ & $-0.58$ & $0.778$ & $0.022$\\
V*XAri & $500.5$ & $[495.3,505.8]$ & $[490.2,511.1]$ & $-0.52$ & $0.438$ & $0.032$\\
V*XXAnd & $1030.7$ & $[1017.5,1044.1]$ & $[1004.4,1057.6]$ & $0.05$ & $0.476$ & $0.029$\\
V*XXPup & $1225.1$ & $[1215.9,1234.4]$ & $[1206.7,1243.8]$ & $0.16$ & $0.601$ & $0.019$\\
V*XZAps & $1880.7$ & $[1864.6,1897.1]$ & $[1848.5,1913.5]$ & $0.91$ & $0.565$ & $0.031$\\
\enddata
\tablenotetext{\fromfit}{Best distance posteriors from the analysis described in \S\ref{plrs}.}
\tablenotetext{\ddesc}{The number of $\sigma$ discrepancy between the prior and posterior mean, defined as $\Delta({\mu}_{\rm prior}-{\mu}_{\rm post}) = \frac{\bar{\mu}_{\rm prior}-\bar{\mu}_{\rm post}}{\sqrt{\sigma_{\mu_{\rm prior}}^2 + \sigma_{\mu_{\rm post}}^2}}$.  }
\tablenotetext{\calcm}{Absolute $V$-band magnitudes calculated with posterior $\mu$ from the PLR analysis (\S\ref{plrs}) and $m^*_V$ from converted, extinction corrected \hipp\ mean-magnitude, (\S\ref{meth}; Table \ref{inputs_table}). }
\end{deluxetable}

\end{document}